\documentclass[aps,prb,showpacs,twocolumn]{revtex4}
\usepackage{epsfig}
\usepackage{amsfonts}
\usepackage{amsmath}
\usepackage[T1]{fontenc}

\newcommand{\beq}{\begin{equation}}
\newcommand{\eeq}{\end{equation}}
\newcommand{\bea}{\begin{eqnarray}}
\newcommand{\eea}{\end{eqnarray}}
\newcommand{\nn}{\nonumber}

\newcommand{\eps}{\epsilon}

\newcommand{\al}{\alpha}
\newcommand{\s}{\sigma}

\newcommand{\be}{\beta}

\newcommand{\p}{\partial}

\newcommand{\ra}{\rangle}
\newcommand{\la}{\langle}

\newcommand{\ga}{\gamma}

\newcommand{\bfs}{\boldsymbol}

\begin{document}
\title{Magnetic Ordering of Nuclear Spins in an Interacting 2D Electron Gas}
\author{Pascal Simon$^{1,2}$, Bernd Braunecker$^1$, and Daniel Loss$^1$}
\affiliation{$^{1}$Department of Physics, University of
Basel, Klingelbergstrasse 82, CH-4056 Basel, Switzerland\\
$^{2}$Laboratoire de Physique et Mod\'elisation des  Milieux
Condens\'es, CNRS and Universit\'e Joseph Fourier, BP 166, 38042
Grenoble, France}

\date{\today}
\begin{abstract}
We investigate the magnetic behavior of 
 nuclear spins embedded in a 2D interacting 
electron gas using a Kondo lattice model description.
We derive an effective magnetic Hamiltonian for the nuclear spins which is of
the RKKY type and where the interactions between the nuclear spins
 are strongly modified by the 
electron-electron interactions.
We show that the  nuclear magnetic ordering at finite temperature relies on the (anomalous) 
behavior of the 2D  static electron spin susceptibility, and thus provides a
connection between 
low-dimensional magnetism and non-analyticities in interacting 2D electron systems.
Using various perturbative and non-perturbative approximation schemes in order to establish the general
shape of the electron spin susceptibility as function of its wave vector, we show that the nuclear spins  locally order 
 ferromagnetically, and that this ordering can become global in certain regimes of interest. 
We demonstrate that the associated Curie temperature for the nuclear system 
increases with the electron-electron interactions
up to the millikelvin range.
\end{abstract}


\maketitle

\section{Introduction}

In the last decade, the field of spintronics has seen
remarkable progress.\cite{awschalom:2002,Cerletti,Hanson}
Among them, the possibility of confining electron spins in quantum dots opens
the door to quantum spintronics. This is based on the
possibility of controlling and manipulating single electron spins
in order to build devices able to achieve operations for quantum information processing.
The most promising and challenging idea is the use of spins of confined electrons in quantum dots
to realize quantum bits.\cite{loss:1998}
Within the last years,
all the necessary requirements for spin-based quantum computation have been realized experimentally,
going from the coherent exchange of two electron spins in a double dot \cite{petta:2005}
to the coherent control of a single electron spin, including the observation of 
Rabi oscillations.\cite{koppens:2006} These achievements have become
possible because electron spins in semiconductor quantum dots are  
relatively weakly coupled to their environment and therefore long lived quantities,
quite robust against decay. Indeed, longitudinal
relaxation times in these systems have been measured to be 
of the order of $1 \mathrm{sec}$.\cite{kroutvar:2004,elzerman:2004,amasha:2006}
A lower bound on the spin decoherence time for an ensemble of electron spins in GaAs quantum dots
has been measured to be typical  larger than  $100\,\mathrm{ns}$,\cite{kikkawa:1998}
while a coherence time in a single quantum dot exceeding 1 $\mu s$ has been 
recently achieved using spin-echo techniques.\cite{petta:2005}
It is by now well established that one of the major sources of decoherence 
for a single electron spin confined
in a quantum dot is the contact hyperfine
interaction with the surrounding lattice nuclear spins.\cite{burkard:1999a}

One possibility to lift this source of decoherence is the development of  quantum control techniques
which effectively lessen or even suppress the nuclear spin coupling to the 
electron spin.\cite{johnson:2005,petta:2005,laird:2006} 
Another possibility is to narrow the nuclear spin distribution, \cite{coish:2004a,Klauser,Stepanenko}
or dynamically polarize the nuclear 
spins.\cite{burkard:1999a,khaetskii:2003,imamoglu:2003,bracker:2005a,coish:2004a} 
However, in order to 
extend the spin decay time  by one order of magnitude through polarization 
of the nuclear spins, a polarization of  above 99\% is required, \cite{coish:2004a}
quite far from 
the best result so far reached in quantum dots, which is around 60\%.\cite{bracker:2005a}
A common point to the aforementioned approaches is their aim at mitigating nuclear spin fluctuations
by external actions. Recently, the possibility was raised of 
an intrinsic polarization of nuclear spins at finite but low temperature
in the two dimensional electron gas (2DEG) confined by the GaAs heterostructure.\cite{simon:2007}

The nuclear spins within the 2DEG interact mainly
via the Rudermann-Kittel-Kasuya-Yosida (RKKY) interaction,\cite{RKKY} which is
mediated by the conduction electrons (the direct dipolar interactions between the nuclear spins are
much weaker, see below).
An intrinsic nuclear spin polarization relies on the existence of a temperature dependent
magnetic phase transition, at which a ferromagnetic ordering sets in,
thus defining a nuclear spin Curie temperature.

The first estimate of such a Curie temperature was obtained for three-dimensional
(3D) metallic samples,
using a Weiss mean field treatment by Fr\"{o}hlich and Nabarro more than sixty years ago.\cite{FN}
They determined the nuclear spin Curie temperature to be
in the microkelvin range or less for 3D metals. A Weiss mean field treatment
also gives a nuclear spin Curie temperature $T_c$ in the microkelvin range for a typical 2DEG
made from GaAs heterostructures,\cite{simon:2007} yet a more detailed analysis is desirable for
at least two reasons. First, a Weiss mean field analysis  does not take into account properly the 
dimensionality of the system, and second ignores electron-electron (e-e) interactions.
In two dimensions (2D), the Mermin-Wagner theorem \cite{MW} states that there is no phase transition
at finite temperature for spin systems with Heisenberg (isotropic) interactions, provided
that the interactions are short-ranged enough.  However, 
RKKY interactions are long-ranged and strictly speaking,
the Mermin-Wagner theorem does not apply,  although
a conjecture extending the Mermin-Wagner theorem
for RKKY interactions due to non-interacting electron systems has been recently formulated 
(and proved in some particular cases).\cite{bruno:2001}

In Ref. \onlinecite{simon:2007}, we started from a Kondo lattice description for the system composed of
nuclear spins and electrons, then derived a rather general effective Hamiltonian for nuclear spins
after integrating out electron degrees of freedom, and finally performed a spin wave analysis around a 
ferromagnetic ground state (which we assumed to be the lowest energy state).
We indeed showed that $T_c=0$ for non-interacting electrons in agreement with the latter conjecture.
However, taking into account e-e interactions changed drastically this conclusion.
It turns out that e-e interactions modify the long range nature of the 2D RKKY
interactions (which are directly related here to the  static electron spin susceptibility) 
and thereby allow some ordering of the nuclear spins at finite temperature.\cite{simon:2007}
Furthermore, we showed that the temperature scale at which this ordering takes place is
enhanced by e-e interactions.\cite{simon:2007}

The study of thermodynamic quantities in {\em interacting}  electron
liquids (especially in 2D) has attracted some 
theoretical \cite{belitz:1997,hirashima:1998,misawa:1998,chitov:2001,maslov:2003,maslov:2006,millis:2006,efetov:2006,shekhter1:2006,shekhter2:2006} and experimental\cite{reznikov:2003}
interest recently  with the goal to find deviations from the standard Landau-Fermi liquid behavior.
It is therefore quite remarkable that the macroscopic  magnetic properties 
of nuclear spins in a 2DEG, and thus their finite temperature ordering, 
are directly related to the corrections to the the  static  electron spin susceptibility 
induced by e-e interactions. They may  therefore be associated with an indirect signature  
of Fermi liquid non-analyticities.
Nevertheless, it turns out that the temperature dependence of the electron spin susceptibility
$\chi_s(T)$ is rather intricate. On the one hand, from perturbative calculations 
in second order in the short-ranged interaction strength one obtains
that $|\chi_s(T)|$ increases with temperature.\cite{maslov:2003,maslov:2006,millis:2006}
The same behavior is reproduced by effective supersymmetric theories.\cite{efetov:2006} On the other hand, 
non-perturbative calculations,
taking into account renormalization effects, found that  $|\chi_s(T)|$ has a non-monotonic behavior
and first decreases with temperature.\cite{shekhter1:2006,shekhter2:2006} 
This latter behavior is in agreement with
recent experiments on 2DEGs.\cite{reznikov:2003} 

In view of these recent controversial results, we want to reconsider the question of a finite
temperature ordering of nuclear spins by taking into account renormalization effects of
the static spin 
susceptibility $\chi_s(q)$, where $q$ is the wave vector, and therefore going beyond 
Ref. \onlinecite{simon:2007}. It turns out that, {\it a priori}, different nuclear spin orderings can occur,
depending on temperature and other sample parameters such as the interaction strength, measured by the
dimensionless parameter 
$r_s$ (essentially, the ratio between Coulomb and kinetic energy of the electrons). We consider at least two possible ordered phases in the nuclear system:
a ferromagnetic ordering\cite{meaculpa}
but also a helical spin ordering where the nuclear spins align ferromagnetically at the scale
of the nuclear lattice constant but point in opposite directions at the scale of the Fermi wave length
(roughly two orders of magnitude larger than the nuclear lattice spacing at small $r_s\lesssim 1$).
Depending on the general
non-perturbative shape of $\chi_s(q)$ (which may have a complex dependence on $T$ and $r_s$),
we discuss the possible ordered phases and their associated magnetic properties.

The outline of the paper is as follows: In Sec. \ref{sec:2}, we formulate a
Kondo lattice description of our problem where the nuclear spins are playing a role analogous to
magnetic impurities embedded in an electron liquid. We then derive a general effective magnetic
Hamiltonian for nuclear spins where the interaction is controlled by the electron spin susceptibility
in 2D. In Sec. \ref{sec:interaction}, we calculate the electron spin susceptibility
in an interacting 2DEG using various approximation schemes for both, short-ranged and long-ranged
interactions. Particular attention is paid to 
renormalization effects in the Cooper channel which turn out to be important. 
Sec. \ref{sec:magnetic} is devoted to the magnetic properties of the nuclear
spins depending on the general wave-vector dependence of the electron spin susceptibility. 
We discuss two different phases: A ferromagnetic phase and a helical phase with a period 
of the order of the electron Fermi wavelength.
Finally, Sec. \ref{sec:conclusion} contains a summary of our main results
and also perspectives. Appendix \ref{sec:SW} contains some details of the derivation
of the effective nuclear spin Hamiltonian and of the reduction to a strictly 2D problem.

\section{Model Hamiltonian}\label{sec:2}

\subsection{Kondo lattice description}
In order to study an interacting electron gas coupled to nuclear spins within the 2DEG, we 
adopt a tight-binding representation in which
each lattice site contains a single nuclear spin and electrons can
hop between  neighboring sites.  A general  Hamiltonian describing such
a system  reads
\bea\label{eq:kl}
H&=&H_0+\frac{1}{2}\sum\limits_{j=1}^{N_l} A_j c^\dag_{j\sigma}{\boldsymbol \tau}_{\sigma\sigma'} c_{j\sigma'}\cdot
{\bf I}_j+\sum\limits_{i,j}
 v_{ij}^{\al\be} { I}_i^\al { I}_j^\be \nn\\
&=&H_0+H_n +H_{dd},
\eea
where $H_0$ denotes the conduction electron 
Hamiltonian, $H_n$ the electron-nuclear spin hyperfine 
interaction and $H_{dd}$ the general dipolar interaction between the nuclear spins.
$H_0$ can be rather general and 
includes electron-electron (e-e) interactions. 
In Eq. (\ref{eq:kl}), $c^\dag_{j\sigma}$ creates an electron at the lattice site 
$\mathbf r_j$ with spin $\sigma=\uparrow,\downarrow$, and $\bfs \tau$ represents the Pauli matrices.
We have also introduced $\mathbf I_j = (I_j^x, I_j^y, I_j^z)$ the
nuclear spin located at the lattice site 
$\mathbf r_j$, and $A_j$ the hyperfine coupling constant between the electron
and the nuclear spin at site $\mathbf  r_j$. 
Summation over the spin components $\al,\beta=x,y,z$ is implied.
The electron spin operator at site $\mathbf  r_j$ is defined by 
$\mathbf S_j= \hbar c^\dag_{j\sigma}\boldsymbol \tau_{\sigma\sigma'}c_{j\sigma'}$
(for convenience we normalize the spin operator here to 1).
$N_l$ denotes the total number of
lattice sites. 
>From here on, we assume that $A_j=A>0$,
which means we assume the hyperfine interaction is antiferromagnetic and  the same for all atoms
that constitute the heterostructures (typically Ga and As and their isotopes).

The nuclear spins are also coupled via the dipolar interaction to other nuclear spins, which are 
not embedded in the 2DEG. Taking into account this interaction as well makes the problem of the magnetism
of nuclear spins in GaAs heterostructures an {\it a priori}
3D tremendously complicated one. Nevertheless, it turns out that the
dipolar interaction energy scale $E_{dd}$ is the smallest 
one. It has been estimated to be $E_{dd}\approx 100~nK$.\cite{paget:1977}
In particular, $k_BT\gg E_{dd}$, where $T$ is the temperature of a typical experiment.
In the rest of the paper, we 
neglect all direct dipolar interactions between the nuclear spins,
which are in general smaller than the indirect interaction, as we will see.
Therefore, we assume that $v_{ij}^{\al\be}\approx 0$ in  Eq. (\ref{eq:kl}).
This assumption is important since it allows us to focus only on those nuclear spins which lie
within the support of the electron envelope wave function (in growth direction).

The general Hamiltonian in Eq. (\ref{eq:kl}) is the well-known Kondo lattice Hamiltonian (KLH),
though $H_0$ contains also e-e interactions. The KLH is one of the most studied models in condensed
matter theory due to its large variety of applications.
The KLH 
has been used to describe the properties of  transition metal oxides,\cite{imada:1998}  
heavy fermions compounds,\cite{lee:1986,sigrist:1997} more
recently also  magnetic semiconductors (or semi-metals) in the series of rare 
earth substances,\cite{nolting}, and  diluted magnetic semiconductors such as
${\rm Ga_{1-x}Mn_xAs}, $\cite{ohno:1998,mcdonald:2006} to list only a few.
The nuclear spins play a role
analogous to magnetic impurities in the Kondo lattice problem. 
The regime in which we are interested corresponds to the weak Kondo coupling regime 
in the sense that $A\ll E_F$,
where $E_F$ is the Fermi energy. Furthermore, the nuclear spin density $n_s$ is far 
larger than the electron density $n_e$.
It is worth noticing that the single nuclear spin Kondo temperature $T_K\approx D\exp(-E_{F}/A)$ (with $D$
being the electron bandwidth)
is extremely small compared to all other energy scales. We are therefore far away from
the so-called controversial exhaustion regime\cite{nozieres} where the individual screening of
the impurity competes with indirect magnetic exchange between the nuclear spins.

In this low electron density regime, 
the ground state of the magnetic system (here the nuclear spins) has been shown to be 
ordered ferromagnetically in 3D 
using various treatments that go beyond mean field theory and which notably 
include spin wave modes (but neglect e-e interactions).\cite{sigrist:1997}

\subsection{Derivation of an effective magnetic Hamiltonian}
We first go to  Fourier space and rewrite $H_n$ in Eq. (\ref{eq:kl}) as
\beq\label{eq:kl1}
H_n=\frac{A}{2N_l}\sum_{\mathbf q} \mathbf S_{\mathbf q}\cdot\mathbf I_{\mathbf q},
\eeq
where
$\mathbf I_{\mathbf q}=\sum_j e^{i \mathbf q\cdot\mathbf r_j}\mathbf I_j$ and
$\mathbf S_{\mathbf q}=\sum_j e^{-i \mathbf q\cdot\mathbf r_j}\mathbf S_j$
are the Fourier transforms of $\mathbf I_j$ and $\mathbf S_j$, respectively.  
(From now on we set $\hbar=1$.)
Since $A$ is a small energy scale in our case, we can perform a Schrieffer-Wolff (SW) transformation
in order to eliminate terms linear in $A$, followed by integrating out the electron
degrees of freedom. Furthermore, we can reduce our initial 3D  model to a genuine
2D problem. The main steps of these calculations are given in Appendix \ref{sec:SW}.
We are left with an effective Hamiltonian $H_{\rm eff}$ for  the nuclear spins
in a 2D plane:
\beq\label{eq:ueff}
H_{\rm eff}=\frac{A^2}{8n_s}\frac{1}{N}\sum\limits_{\mathbf q} I_{\mathbf q}^\al~
\chi_{\al \be}( q) ~I^{\be}_{-\mathbf q}~,
\eeq
where
\beq
\chi_{\al\be}(\mathbf q,\omega)=-\frac{i}{Na^2}\int_0^\infty dt~ e^{-i\omega t-\eta t}
\la[ S_{\mathbf q}^\al(t),S_{-\mathbf q}^\beta]\ra,
\eeq
is a general 2D electron spin susceptibility tensor,
$\chi_{\al\be}(q) = \chi_{\al\be}(\mathbf q,\omega=0)$ and $q=|\mathbf{q}|$.
$N$ is the number of lattice sites in the 2D plane and $a$ denotes the lattice 
spacing for nuclear spins. 
Note that $\la\dots\ra$ means average over electron degrees of freedom only.
We have normalized $\chi$ such that it coincides with the density-density Lindhard function
(see below) in the isotropic and non-interacting limit.

The only assumptions we make are 
time reversal symmetry of $H_0$, as well as
translational and rotational invariance. The effective Hamiltonian in Eq. (\ref{eq:ueff}) 
is therefore
quite general and does not depend on the dimensionality of the system.
Note that Eq. (\ref{eq:ueff}) is also valid
when electron-electron interactions are taken into account. 
It is worth emphasizing that the SW transformation neglects retardation effects.
This is appropriate since the
the nuclear spin dynamics is  slow compared to 
the electron one (in terms of energy scales this is related to the fact that $A\ll E_F$). 
Therefore, electrons see an almost static nuclear spin background, and the adiabatic
approximation (for the conduction electrons) is well justified. 
In the case of ferromagnetic semiconductors, such an approximation breaks
down and retardation effects must be taken into account.\cite{konig:2000}
If we also assume spin isotropy in the 2DEG, then 
$\chi_{\al \be}(q,\omega\to 0)=\delta_{\al \beta} \chi_{s}(q)$, where $\chi_{s}(q)$ 
is the isotropic electron spin susceptibility in the static limit.

In real space, the effective nuclear spin Hamiltonian reads
\beq\label{eq:hreal}
H_{\rm eff}=-\frac{1}{2}\sum_{\mathbf r,\mathbf r'} J_{\mathbf r-\mathbf r'}^{\al\be}
 I_{\mathbf r}^\al I_{\mathbf r'}^\be,
\eeq
where 
\beq\label{eq:J}
 J_{|\mathbf r|}^{\al\be}=-({A^2}/{4n_s})\chi_{\al\be}(|\mathbf r|),
\eeq
 is the effective exchange coupling.
The nuclear spins $\mathbf {I}_{\mathbf r}$
are therefore interacting with each other, this interaction being mediated by the conduction electrons.
This is just the standard RKKY interaction,\cite{RKKY} 
which, however, as we shall
see, can be substantially modified by electron-electron interactions compared to the free electron case.

\section{Electron spin susceptibility in a 2D interacting electron gas}\label{sec:interaction}

The main result of the previous section is that the magnetic exchange interaction between the nuclear spins 
is mediated by the electron gas. Therefore, the key quantity governing the magnetic properties
of the nuclear spins is the electron spin susceptibility $\chi_s(\mathbf q)$ in two dimensions.
The calculation of this quantity in an interacting 2DEG has been the subject of intense
efforts in the last decade in connection with non-analyticities in the Fermi 
liquid theory.\cite{belitz:1997,hirashima:1998,misawa:1998,chitov:2001,maslov:2003,maslov:2006,millis:2006,efetov:2006,shekhter1:2006,shekhter2:2006}
On a more fundamental level, incorporating e-e 
interactions in the calculations of thermodynamic quantities has been 
an important area of condensed matter theory over the last fifty years. In particular, 
the study of non-analytic behavior of thermodynamic quantities and susceptibilities in electron liquids has attracted recent interest, 
especially in 2D.\cite{belitz:1997,hirashima:1998,misawa:1998,chitov:2001,maslov:2003,maslov:2006,millis:2006,efetov:2006,shekhter1:2006,shekhter2:2006} 
Of particular importance for the following is the recent findings 
 by Chubukov and Maslov \cite{maslov:2003}, namely that the static non-uniform spin susceptibility 
$\chi_s(q)$ depends {\em linearly } on the wave vector modulus $q=|\mathbf q|$ for $q \ll k_F$ in 2D (while it is $q^2$
in 3D), with $k_F$ the Fermi momentum.  This non-analyticity arises from the long-range correlations between quasi-particles mediated by 
virtual particle-hole pairs, despite the fact that e-e interactions was assumed
to be short-ranged.

Let us first recall the case of non-interacting electrons. In this case,
$\chi_{s}$ coincides with the usual  density-density (or Lindhard) response 
function $\chi_{L}$,\cite{GV}
\beq
\chi_{L}(q)=\frac{1}{Na^2}\sum\limits_{\mathbf k,\s}\frac{n_{\mathbf k,\s}-n_{\mathbf k+\mathbf q,\s}}
{\epsilon_{\mathbf k,\s}-\epsilon_{\mathbf k+\mathbf q,\s}+i\eta},
\eeq
where $n_{\mathbf k}$ is the electron number operator, $\eps_{\mathbf k}$ the dispersion relation,
and $\eta>0$ is an infinitesimal regularization parameter.
This Lindhard function can be evaluated exactly and reads in 2D \cite{GV}
\beq\label{eq:lindhard}
\chi_L(q)=-N_e\left(1-\Theta(q-2k_F)\frac{\sqrt{q^2-4k_F^2}}{q}\right),
\eeq
where $N_e=n_e/E_F$ is the electron 
density of states (per spin).  
Note that $N_e= m^*/\pi$ where $m^*$ is the effective electron mass in a 2DEG.
It follows from Eq. (\ref{eq:lindhard})  that
\beq
\delta\chi_L(q)\equiv\chi_L(q)-\chi_L(0)=0 ~~{\rm for}~~q\leq 2k_F.
\eeq

Let us now include electron-electron interactions.
It is convenient to introduce a relativistic notation with $\bar p\equiv (p_0,\mathbf p)$ 
being the  (D+1)-momentum
where $p_0$ denotes the frequency and $\mathbf p$ the $D$-dimensional  wave vector (here $D=2$).
In a zero-temperature formalism, the susceptibility can be written diagrammatically\cite{GV} as

\beq\label{eq:chis_def}
\chi_s(\bar q)=-\frac{i}{L^D}\sum\limits_{\bar p_1,\s,\s'} 
\s\s' G_\s(\bar p_1-\bar q/2)G_\s(\bar p_1+\bar q/2)\Lambda_{\bar p_1\s\s'}(\bar q), 
\eeq
with $\sigma,\sigma'=\pm$, and 
where $L=a N^{1/D}$ is the system length, $G_\s(\bar p_1)$ 
is the exact single-particle Green's function 
and $\Lambda(\bar q)$ is the exact vertex function, which can be 
expressed in terms of the exact 
scattering amplitude $\Gamma(\bar q)$ as follows:\cite{GV}
\bea
&&\Lambda_{\bar p_1\s\s'}(q)=\delta_{\s\s'}~-\\
&&\frac{i}{L^{D}}\sum\limits_{\bar p_2}\Gamma_{\bar p_1,\bar p_2}^{\s\s\s'\s'}(\bar q) 
G_{\s'}(\bar p_2+\bar q/2)G_{\s'}(\bar p_2-\bar q/2).\nn
\eea
This scattering amplitude plays a crucial role as we see next. $\Gamma$ is for a 
general scattering event a function of four spin variables $\s_1,\s_1',\s_2,\s_2'$. Nevertheless, 
one can use a convenient parametrization which ensures rotational spin invariance,\cite{GV}

\beq
\Gamma_{\bar p_1,\bar p_2}^{\s_1\s_1'\s_2\s_2'}(\bar q)
=\Gamma_{\bar p_1,\bar p_2}^+\delta_{\s_1\s_1'}\delta_{\s_2\s_2'}
+\Gamma_{\bar p_1,\bar p_2}^-\bfs\tau_{\s_1\s_1'}\cdot\bfs\tau_{\s_2\s_2'},
\eeq   
where $\bfs \tau$ is a vector whose components are the Pauli matrices $(\tau^x,\tau^y,\tau^z)$
and $\bfs\tau_{\s_1\s_1'}\cdot\bfs\tau_{\s_2\s_2'}=\sum\limits_{a=x,y,z}\tau^a_{\s_1\s_1'}\tau^a_{\s_2\s_2'}$.
Note that
$\Gamma^{\pm}$ is spin-independent and corresponds to the charge and spin channels, respectively.
Following Ref. \onlinecite{GV}, we next write the 
Bethe-Salpeter (BS) equation for $\Gamma^-$ (corresponding to the spin channel) as follows:
\bea \label{eq:BS}
&&\Gamma^-_{\bar p_1\bar p_2}(\bar q)=(\Gamma_{\rm irr}^-)_{\bar p_1\bar p_2}(\bar q)
+\\
&&\frac{1}{L^D}\sum\limits_{\bar p''}(\Gamma^-_{\rm irr})_{\bar p_1\bar p''}
(\bar q)R_{\bar p''}(\bar q)\Gamma^-_{\bar p''\bar p_2}(\bar q),\nn
\eea
where $ (\Gamma^-_{\rm irr})_{\bar p\bar p'}(\bar q)$ is the exact irreducible electron-hole
 scattering amplitude in the spin channel, and 
\beq
 R_{\bar p}(\bar q)=-2i G(\bar p+\bar q/2)G(\bar p-\bar q/2)
\eeq
 is the electron-hole bubble.\cite{note_GV}

One can exactly solve, at least formally, the
BS equation (\ref{eq:BS}) using a matrix notation where
the matrix indices run over $\bar p$. Within this notation $R$ is a diagonal matrix.
We find that
\beq
\Gamma^-_{\bar p_1\bar p_2}=\sum_{\bar p''} (\Gamma^-_{\rm irr})_{\bar p_1\bar p''}
\left(\frac{1}{1-\Gamma_{\rm irr}^-(\bar q)R(\bar q)} \right)_{\bar p''\bar p_2}.
\eeq
This enables us to  derive an exact and closed expression for the spin susceptibility, given by
\beq\label{eq:chis}
\chi_s(\bar q)=\frac{1}{L^{D}}\sum\limits_{\bar p,\bar p'}
\left( R(\bar q)\frac{1}{1-\Gamma_{\rm irr}^-(\bar q)R(\bar q)}\right)_{\bar p\bar p'}\, .
\eeq

In general, $\Gamma^-_{\rm irr}$ cannot be calculated exactly and some approximations are required.
The approximation we use in the following consists in replacing the
exact irreducible electron-hole scattering amplitude
$(\Gamma^-_{irr})_{\bar p,\bar p'}$  by an averaged value calculated 
with respect to all possible values 
of $p$ and $p'$
near the Fermi surface. This is equivalent to the following approximation: 
\beq
(\Gamma^-_{\rm irr})_{\bar p,\bar p'}(\bar q) \approx \Gamma^-_{\rm irr}(\bar q)~\forall~p,p'.
\eeq 

We now assume $q_0=0$ and suppress the $q_0$-argument in what follows since we are interested in
the static properties of the spin susceptibility.

\subsection{Short-ranged interaction}

In this section, we consider 
a $q$-independent short-ranged interaction potential, 
which corresponds within our notations to $\Gamma^-_{\rm irr}(q)=-U$. 
This approximation considerably simplifies the BS equation (\ref{eq:BS}) and the formal
expression of $\chi_s$ in Eq. (\ref{eq:chis}). 
The derivative of $\chi_s(q)$ with respect to $q$ can be expressed in a simple compact form:
\beq\label{eq:deltachis}
\frac{\partial\chi_s}{\partial q}(q)
=\frac{\partial\Pi(q)}{\partial q}\frac{1}{(1+U\Pi(q))^2},
\eeq
where $\Pi(q)=\sum_{\bar p} R_{\bar p}(q)/L^{D}$. In the low $q\ll k_F$ limit, one can approximate the term 
$\Pi(q)$ in the denominator of Eq. (\ref{eq:deltachis})
by its non-interacting value  $\chi_L(0)=-N_e$. 

The multiplicative  factor $1/(1-UN_e)^2$ in Eq. (\ref{eq:deltachis}) 
signals the onset of 
the ferromagnetic Stoner instability
when $UN_e$ approaches unity. The Stoner instability is supposed to occur
for very large $r_s\sim 20$ according to Monte Carlo results.\cite{montecarlo}
For smaller $r_s\leq 10$, we are still far from the Stoner instability. Though this multiplicative term
does not play a significant role at small $r_s$, it increases with $r_s$, showing the tendency.

\subsubsection{Perturbative calculation}

The corrections to the polarization bubble $\Pi(q)$ are 
dominated by the first bubble correction
to the self-energy of $G_\s(p)$.\cite{maslov:2003} These corrections 
have been calculated in second order
in $U$ in the small $q$ limit in Ref. \onlinecite{maslov:2003},
with the result 
\beq\label{eq:pert}
\delta\Pi(q)=\Pi(q)-\Pi(0)\approx -4q \chi_{S}\frac{|\Gamma(\pi)|^2}{3\pi k_F},
\eeq
where $\chi_S = |\chi_{s}(0)|$ and 
$\Gamma(\pi)\sim-Um^*/4\pi$ 
is the $(2k_F)$ backscattering amplitude. 
When $U N_e\ll 1$, we recover from Eq. (\ref{eq:deltachis}) the known result 
$\delta \chi_s(q)=\delta \Pi(q)$.\cite{maslov:2003}  
This perturbative calculation therefore gives that $\delta \chi_s(q)= \alpha q$ with $\alpha<0$, 
or equivalently that $|\chi_s(q)|$ {\it increases} with $q$ at low $q$.\cite{meaculpa}
Finite temperature calculations along the same line imply that
$\delta \chi_s(T)= \gamma T$ with $\gamma<0$.\cite{maslov:2006,millis:2006}
This result has been confirmed by using a supersymmetric effective theory 
of interacting spin excitations.\cite{efetov:2006}

On the other hand, recent experiments on a low density interacting electrons gas in Silicon MOSFETs  
have found
that $|\chi_s(T)|$ decreases at low temperature\cite{reznikov:2003} 
in apparent contradiction with perturbative calculations. This rather puzzling situation
demands a non-perturbative approach.

\subsubsection{Beyond lowest order perturbation theory: renormalization effects}\label{sec:RPTA}

The previous calculations give a spin susceptibility which is quadratic in the backscattering amplitude.
However, it is known that at low enough energy the backscattering amplitude becomes renormalized.
One should therefore instead consider some type of renormalized perturbation theory approximation (RPTA).
Shekhter and Finkel'stein \cite{shekhter1:2006} argued recently that the strong 
renormalization of 
the scattering amplitude in the Cooper channel
may  explain the sign of $\delta \chi_s(T)$ in the experiment by Prus {\it et al.}\cite{reznikov:2003}

Let us introduce $\Gamma(\theta,T)$,
the two-particle scattering amplitude at a particular scattering angle 
$\theta$ and temperature $T$. The Cooper channel corresponds to $\theta=\pi$
and we denote by $\Gamma_c(T)=\Gamma(\pi,T)$ the corresponding two-particle scattering amplitude.
In the Cooper channel the two particles that scatter have exactly opposite momenta.
We first expand $\Gamma_c$ in Fourier harmonics such that 
\beq\Gamma_c (T)=\sum_n (-1)^n \Gamma_{c,n}(T),
\eeq
where $n$ is an integer representing the angular momentum quantum.

In order to analyze the temperature dependence of  $\Gamma_{c,n}(T)$, 
it is enough
to write the BS equation for $\Gamma_c$ in a way fully analogous to Eq. (\ref{eq:BS}).
In the Cooper channel, due to the fact that the momentum of the center of motion of the
two scattering particles is zero, the integration over the electron-hole bubble 
gives rise to
logarithmic infrared (IR) divergences. These divergences are 
due to particle-hole excitations around the Fermi sea and already occur 
in the ladder approximation ({\it i.e.} by considering only the bare short-ranged interaction
potential in the BS equation).
We refer the reader to references \onlinecite{saraga:2005} and \onlinecite{shekhter1:2006}
for more details. The logarithmic IR divergences can be absorbed by a
rescaling of the scattering amplitudes in the Cooper channel such that:
\beq\label{eq:gamma_r}
\Gamma_{c,n}(T)= \frac{\Gamma_{c,n}}{1+\Gamma_{c,n}\ln(E_F/k_BT)},
\eeq
where $\Gamma_{c,n}=\Gamma_{c,n}(E_F)$ is the bare value. 
The logarithmic factors give rise to the well known  Cooper instability.\cite{cooper:1956,mahan}
They are just the 2D equivalent of the one found in the discussion
of Kohn-Luttinger superconductivity.\cite{kohn:1965}
In Eq. (\ref{eq:gamma_r}), the logarithmic divergence is cut-off by the temperature at low energy.

Another way of recovering Eq. (\ref{eq:gamma_r}) is to regard
the backscattering amplitudes $\Gamma_{c,n}$  as some energy
scale dependent coupling constants $\Gamma_{c,n}(\Lambda)$
of the 2DEG ($\Lambda$ is some running energy scale, the equivalent of $T$ in Eq. (\ref{eq:gamma_r})). The next step is to 
write renormalization group equations (RG) for these couplings.
These couplings are marginal and the RG equations at lowest order read:
\beq
\frac{d\Gamma_{c,n}}{d\ln(\Lambda/E_F)}=-(\Gamma_{c,n})^2.
\eeq
The solutions of the RG
equations for $\Gamma_{c,n}(\Lambda)$ are given 
by Eq. (\ref{eq:gamma_r}) where essentially $T$ is replaced by $\Lambda$.\cite{shekhter1:2006}

The low energy behavior of the $\Gamma_{c,n}(T)$ strongly depends 
on the bare scattering amplitudes
being repulsive ($\Gamma_{c,n}>0$) or attractive  ($\Gamma_{c,n}<0$). 
>From Eq. (\ref{eq:gamma_r}), we immediately infer that when $\Gamma_{c,n}>0$, $\Gamma_{c,n}(T)\to 0$ 
at low energy/temperature. On the
other hand, when  $\Gamma_{c,n}<0$, $\Gamma_{c,n}(T)$ renormalizes to the strong coupling regime.
Assuming there exists at least one harmonic $n_0$
such that $\Gamma_{c,n_0}<0$ implies that there
exists a temperature below which the scattering in the Cooper channel is entirely dominated by
$\Gamma_{c,n_0}$.\cite{shekhter1:2006}

This reasoning relies on the fact that at least one 
bare scattering amplitude in the $n^{th}$ harmonic
is negative. This assumption can be further substantiated with some explicit 
perturbative calculations of 
the irreducible scattering amplitudes.\cite{kohn:1965,chubukov:1993,galitski:2003}
By  computing the lowest order corrections contributing to the irreducible 
scattering amplitude $\Gamma_{\rm c,irr}$, 
(therefore going beyond the ladder approximation in the BS equation which leads to Eq. (\ref{eq:gamma_r}))
one can actually prove that there exist higher harmonics 
such that $\Gamma_{c,n}<0$.\cite{saraga:2005}

Note that these calculations imply that a superconducting  instability 
should develop
at a temperature $T\lesssim T_{L}= E_F\exp(-1/|\Gamma_{c,n_0}|)/k_B$ . 
This is entirely analogous to the Kohn-Luttinger mechanism for 
superconductivity.\cite{kohn:1965,luttinger:1966} 
Nevertheless, in a typical 2DEG, disorder 
or some other mechanism may provide 
a natural infrared cut-off preventing this superconducting
instability to be reached. Let us call $\Delta>T_L$ this infrared cut-off.
It is worth noting that $\Delta$ should not be too large either in order to let
$\Gamma_{c,n_0}$ flow to relative large values of order one for this mechanism to be relevant.

As shown in Ref. \onlinecite{shekhter1:2006} this mechanism, if it takes place, 
leads to a non-monotonic
behavior of the temperature dependence of the electron spin susceptibility and more specifically
to the existence of a scale $T_0$, below which $\frac{d\chi_s}{dT}>0$ (note that we use a 
different sign convention for $\chi_s(0)$ compared to Ref. \onlinecite{shekhter1:2006}).
By fitting the experimental data of Prus {\it et al.} with such a theory,   the estimate
$\Gamma_{c,n_0}\sim 0.25-0.3$ was obtained in Ref. \onlinecite{shekhter1:2006}, which
implies $T_0\sim 10~K$, a surprisingly large temperature scale of the order
$E_F\sim 40~K$ in this experiment. It is worth emphasizing that such a scale, being dependent
on the bare value of $\Gamma_{c,n_0}$, is non-universal and therefore sample dependent.

This RPTA also raises a similar issue concerning the  behavior of the electrostatic 
spin susceptibility $\chi_s(q)$ and more specifically of $\Gamma_{c}(q)$.
Following Ref. \onlinecite{saraga:2005}, one can write  a BS equation at zero temperature 
for  $\Gamma_{c}(\xi)$, where
$\xi$ is an energy scale in the vicinity of the Fermi energy $E_F$.
By linearizing $\xi$ around $E_F$ such that $\xi= v_F q$, one immediately infers from 
(\ref{eq:gamma_r}) that
\beq\label{eq:gamma_r1}
\Gamma_{c,n}(q)= \frac{\Gamma_{c,n}}{1+\Gamma_{c,n}\ln(E_F/v_Fq)}\approx 
 \frac{\Gamma_{c,n}}{1+\Gamma_{c,n}\ln(k_F/2q)}.
\eeq
Note that at finite temperature and for a realistic system, the infrared cut-off $\Lambda$ is replaced
by $\max\{v_Fq,k_BT,\Delta\}$.
This result is also obtained from a renormalization group approach by 
working directly in momentum
space at low energy where the energy can be linearized around $k_F$. 
At low enough momentum, $\Gamma_c(\theta,q)$ is dominated by the  harmonic $n_0$
such that
\beq\label{eq:gamma0}
\Gamma_c(q)\approx (-1)^{n_0}\frac{-|\Gamma_{c,n_0}|}{1-|\Gamma_{c,n_0}|\ln(k_F/2q)}.
\eeq
By replacing the bare value of the scattering amplitude in Eq. (\ref{eq:pert}) 
by the renormalized one, 
we find that
$\delta \chi_s(q)\approx -4 q N_e\frac{|\Gamma(\pi,q)|^2}{3\pi k_F}$ also acquires a non-trivial 
$q$-dependence. In particular, using Eq. (\ref{eq:gamma0}), we obtain that
\beq\label{eq:chi_ren}
\frac{d\chi_s}{dq}\approx -\frac{4 N_e}{3 \pi k_F}(\Gamma_c^2-2|\Gamma_c^3|),
\eeq
which is positive when $|\Gamma_c(q)|>1/2$.

Let us assume that $\Delta$, the aforementioned infrared cut-off, is close
to the Cooper instability, {\it i.e.} $k_BT_L\lesssim \Delta$.
In such a case, the flow of $\Gamma_c(q)$ in Eq. (\ref{eq:gamma0}) is cut off by $\Delta$.
Using $1/\Gamma_{c,n_0}=\ln(E_F/k_BT_L)$, one obtains that
\beq \label{eq:gamma_cutoff}
\Gamma_c\approx \frac{(-1)^{n_0+1}}{\ln(\frac{\Delta}{k_BT_L})}~~{\rm for}~q\lesssim \Delta/v_F
\eeq
is renormalized toward large values. This implies that there exists a 
$q_0\sim 10 q_{L}$ (where $q_L=k_BT_L/v_F$) 
such that $\frac{d}{dq}\delta \chi_s(q)>0,~\forall~q<q_0$, and
$\frac{d}{dq}\delta \chi_s(q) < 0 $ when $q \gtrsim q_0$. 

We must point out that the RG equation in Eq. (\ref{eq:chi_ren}) assumes that all other
scattering amplitudes $\Gamma_n$ for $n\ne n_0$ can be neglected below some low energy. 
In order to  go beyond this approximation, one has to solve  (numerically)
a set of RG differential equations instead.

This scenario relies on a fine tuning of $\Delta$ compared to $k_BT_L$ (typically
this demands $T_L\lesssim\Delta/k_B<10 T_L$). If this condition is not met, one may then expect 
$\delta\chi_s(q)<0$ in accordance with lowest order perturbative calculations.
Nevertheless, one should mention that an alternative theory, also giving $\delta\chi_s(T)>0$
at low $T$, has been put forward.\cite{shekhter2:2006}
This alternative scenario applies for vanishing Cooper amplitudes. In such a case,
the anomalous temperature dependence of the spin susceptibility is dominated by
non-analytic contributions from particle-hole rescattering with small momentum transfer.\cite{shekhter2:2006}
Whether such a scenario, not considered here, also implies that $\delta\chi_s(q)>0$ is 
an interesting but open question.

The aforementioned considerations immediately raise the issue about the amplitude of $q_0$ 
in a typical interacting 2DEG. 
In order to describe  the temperature dependence of $\chi_s(T)$, Shekhter and Finkel'stein
determined $\Gamma_{n_0}$ such that the experimental behavior of Ref. \onlinecite{reznikov:2003} 
is reproduced. This fixes the value
of this parameter and also the scale $k_B T_0$ to the order of $E_F$. One may therefore expect $q_0$ to be of 
the order of $k_F$ for a similar 2DEG.
Another independent way of substantiating these estimates is to go beyond the ladder approximation
in the BS equation.\cite{saraga:2005} 
First, as we mentioned before, this allows one to show that there
exists a value $n_0$ at which $\Gamma_{c,n_0}<0$. Second, this can give us an estimate
for $T_L$ and therefore for $q_0$. 
Indeed, in Ref. \onlinecite{saraga:2005} the Cooper instability has been estimated to set in
at the temperature $k_BT_L\sim E_Fe^{-16E_F/(r_s^3\xi)}$,
where $\xi\sim\Delta$ plays the role of the infrared cut-off in Ref. \onlinecite{saraga:2005},
and the condition  $\xi/E_F\ll 1$ was assumed.
Consistency requires that $T_L\sim \Delta \sim E_F$ at large $r_s$, implying 
$q_0\sim k_F$. This is in agreement
with our  previous estimate.

\begin{figure}[h]
\hskip -1.cm \psfig{figure=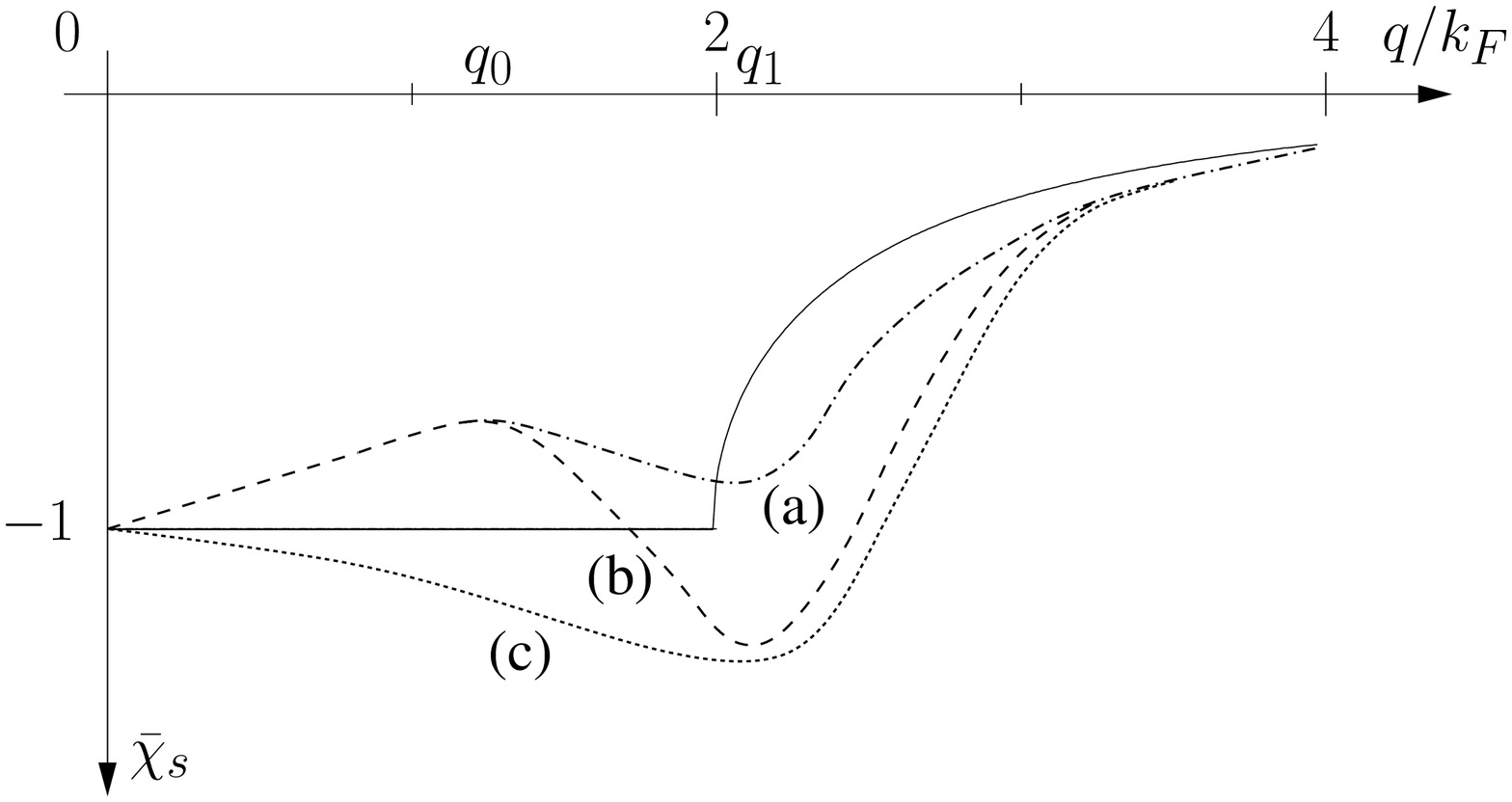,width=7.6cm}
\caption{Three possible qualitative shapes denoted by (a), (b) and (c) 
for the normalized susceptibility $\bar\chi_s(q)=\chi_s(q)/|\chi_s(0)|$ 
as a function of $q/k_F$ (dashed lines) compared to the non-interacting value (thin full line).
Here, $q_0$ and $q_1$ are the positions of the extrema for curves (a) and (b).
In contrast, the local field factor approximation discussed in Sec. \ref{sec:lffa}
results in a monotonic increase of $\bar \chi_s(q)$ (not sketched in the figure), being
always larger than the non-interacting value.}
\label{Fig:chis}
\end{figure}

The previous calculations also give us information about the possible shapes for
 $\chi_s(q)$. 
When renormalization in the Cooper channel is important, we obtain at least one extremum 
around some wave vector $q_0\sim O(k_F)$. Furthermore, at large $q$, we 
should recover the non-interacting behavior and therefore  $\chi_s(q)\sim \chi_L(q)\to 0$
for $q\to \infty$.
Because $\chi_s(q_0)<0$, we 
expect another extremum around a value $q_1>q_0$. Since the non-interacting behavior  is
recovered for $q\gg 2k_F$, one may suspect $q_1\sim O(2k_F)$. From the previous considerations
we therefore conclude that there exist 
(at least) two extrema for the electron spin susceptibility $\chi_s(q)$.
It is worth emphasizing that this double-extremum structure is a direct consequence of the
nontrivial renormalization of the scattering amplitude in the Cooper channel.
We have schematically drawn in Fig. \ref{Fig:chis} the possible qualitative shapes denoted by 
(a) and (b) of $\bar\chi_s(q)=\chi_s(q)/|\chi_s(0)|$ as a function of $q/k_F$ 
and compared it to the (normalized) 
non-interacting $\chi_L(q)/N_e$ at $T=0$. In the case denoted by (a), we choose $\chi_s(q_2)>\chi_s(0)$,
whereas in the case denoted by (b) 
the absolute value of the susceptibility at $q_2 \simeq 2k_F$ is chosen to exceed
the static value, {\it i.e.} $\chi_s(q_2)<\chi_s(0)$.
The previous considerations do not allow us to discriminate between these two possible shapes
of $\chi_s(q)$. Furthermore, by increasing $r_s$, $\chi_s$ can evolve from one shape 
to another.

On the other hand, if the renormalization in the Cooper channel does not take place, {\it e.g.}
when it is cut off by disorder, then the perturbative calculations at lowest order 
apply and  give instead that $\delta\chi_s(q)<0$ at low $q$.\cite{maslov:2003,maslov:2006,millis:2006}
A possible shape for $\chi_s(q)$,
consistent with these calculations, has been  drawn in Fig.  \ref{Fig:chis} and corresponds to label 
(c). We should note, however, that the effect of rescattering of a pair of quasiparticles
in all different channels \cite{shekhter2:2006} should be carefully examined  and may still
lead to shapes (a) or (b) in Fig. \ref{Fig:chis}.

\subsection{Long-ranged Coulomb interactions} \label{sec:lffa}

In the preceding section, we replaced $\Gamma_{\rm irr}^-(q)$ 
by an almost $q-$independent constant operator,
assuming that the Coulomb interaction was screened and, therefore, short-ranged.
Let us consider in this section the bare 2D Coulomb interaction, 
$V(q)=2\pi e^2/q$,  where $e$ is the electron charge.

\subsubsection{Local Field Factor Approximation}\label{sec:lffa0}

One of the most successful
approximations for the calculation of electron response functions
is the local field factor approximation (LFFA).
It improves the random phase approximation \cite{GV} for which 
the effective field seen by an electron is the field that would be seen by a 
classical test charge embedded in the electron gas.
The idea of the LFFA to correct the  random phase approximation
and to better account for the correlations  existing in the electron gas,
is to replace the average electrostatic potential 
by a local field 
effective potential
seen by an electron with spin $\s$ which is part of the 2DEG. (We refer to Ref. \onlinecite{GV} for a review).
The local field factor $G_-$ can be defined as follows:
\beq
G_-(q)V(q)=\chi_s^{-1}(q)-\chi_L^{-1}(q),
\eeq
or, equivalently, the static spin susceptibility $\chi_s$ can be written as
\beq\label{eq:chi_lfft}
\chi_s(q)=\frac{\chi_L(q)}{1+{V}(q)G_-(q)\chi_L(q)}.
\eeq
The precise determination of $G_-(q)$ for all $q$ is  an open problem. 
However, the asymptotic regimes, particularly the $q\to 0$ limit,
are quite well established because they are strongly constrained by sum rules.\cite{GV}
In this work, we use
a semi-phenomenological interpolation formula given
in Ref. \onlinecite{GV} 
\beq\label{eq:lfft}
G_-(q)\approx g_0\frac{q}{q+g_0(1-\chi_P/\chi_S)^{-1}\kappa_2}.
\eeq
Here,  $(g\mu_B)^{-2}\chi_P$ is the Pauli susceptibility 
($\mu_B$ the Bohr magneton and $\chi_P>0$), $\chi_S=|\chi_s(0)|$ 
the renormalized value of the spin susceptibility
at $q=0$,
$\kappa_2=k_F r_s\sqrt{2}$ is the 2D Thomas-Fermi wave vector, 
and $g_0$ is 
the pair-correlation function at $r=0$, describing the probability of finding two electrons 
(of opposite spins) 
at the same position in the electron gas.
This phenomenological form for $G_-$ has been modified from the one originally 
proposed by Hubbard \cite{hubbard}
in order to satisfy exactly 
the compressibility sum rule.\cite{pines,GV} 
The main weakness of this approach is the  arbitrariness of the chosen form for $G_-$.
For non-interacting electrons $\chi_P/\chi_S=1$.
An approximate form for $g_0$ giving a good agreement with quantum Monte Carlo (QMC) 
calculations has been proposed recently by Gori-Giorgi {\it et al.}
and reads:\cite{gori:2004}
\beq\label{eq:g0}
g_0(r_s)\approx (1+Ar_s+B r_s^2+Cr_s^3)e^{-Dr_s}/2.\eeq
The parameters $A=0.088,~B=0.258,~C=0.00037,~D=1.46$ are fitting parameters reproducing QMC results
for $g_0$ in a 2DEG.\cite{gori:2004}
This approximation yields
\beq\label{eq:chi_lfft1}
\chi_s(q)\approx -N_e\frac{q+g_0\kappa_2\alpha}{q+g_0\kappa_2(\al-1)}~~{\rm for}~~q<2q_F,
\eeq 
where $\al=\chi_S/(\chi_S-\chi_P)$ can be regarded as a Fermi liquid parameter.
The low-$q$ semi-phenomenological approximation for the electron spin susceptibility given in
Eq. (\ref{eq:chi_lfft1}) results in $\frac {d\chi_s(q)}{dq}>0, ~\forall q$, in 
contrast to the lowest order perturbative calculations.\cite{maslov:2003,maslov:2006,maslov:2006}

Note that a direct estimate of $G_-(q)$ by recent QMC in a 2DEG 
gives an almost  linear in $q$ behavior 
up to rather large values of $q\lesssim 2k_F$, followed by a more complex non-monotonic 
behavior around $2k_F$, and finally diverges in the large-$q$ limit.\cite{moroni:1995,qmc,GV}
This large-$q$ limit is not reproduced by Eq. (\ref{eq:lfft}). This is not a serious 
drawback since most quantities of interest are dominated by the low-$q$ regime.

However, the scale $q^*=g_0\kappa_2(\al-1)$ decreases exponentially with $r_s$ according to Eq. (\ref{eq:g0}). 
This would imply
an almost constant behavior for $G_{-}(q)$ except at low $q$.
When we compare this behavior with available QMC data,\cite{qmc,GV} 
we find that there is a manifest contradiction.
Therefore, this raises some doubt about the presence of $g_0$ (a short
distance quantity) in Eq. (\ref{eq:lfft}).

\subsubsection{Modified Local Field Factor Approximation}\label{sec:lffa1}

If we instead replace $g_0$ in Eq. (\ref{eq:lfft})
by a parameter $g_1$, such that $g_1\kappa_2(\al-1)=g_1r_s\sqrt{2}(\al-1)k_F\gg 2k_F$, 
we have checked that the QMC data for $G_-(q)$ are  much better reproduced for q$'s$ up to $2k_F$
than by the expression given in Eq. (\ref{eq:lfft}). 
In such a modified local field factor approximation (MLFFA), $g_1$ is approximately given by
\beq\label{eq:g1} 
g_1\approx \frac{\zeta_1}{r_s(\al-1)}=\frac{\zeta_1}{r_s}\left(\frac{\chi_S}{\chi_P}-1\right),\eeq
with $\zeta_1$ some numerical constant of order one.

We should mention that some other more complicated analytical fits of the QMC data
have been obtained in Ref. \onlinecite{davoudi:2001}. Nevertheless, we note that the fits
used in that paper lead to $\delta\chi_s(q)\sim q^2$ for 2D, 
which is in contradiction with
all previous approximations.
It seems desirable to test Eq. (\ref{eq:g1}) with more detailed QMC calculations.

\subsection{Comparison of the various approximation schemes}

If we summarize the various approximation schemes presented in the previous sections, which are  perturbative or
semi-phenomenological, we can clearly ascertain that $\delta\chi_s(q)\propto q$ for $q\ll q_F$.
Nevertheless, the sign of the proportionality constant depends on the approximation scheme we used.

Lowest order perturbation theory in the interaction strength leads to 
$d\chi_s(q)/dq<0$ at low $q$. \cite{maslov:2003,maslov:2006} However, within the RPTA,
renormalization effects in the Cooper channel\cite{saraga:2005,shekhter1:2006} 
are important and  change
the picture given by lowest order perturbation theory. In this latter case,
the RPTA yields an opposite sign for $d\chi_s(q)/dq$ 
below some wave vector $q_0$.
The LFFA we used 
implies that $d\chi_s(q)/dq>0$ for all $q$ and therefore a monotonic behavior (not shown in Fig.  \ref{Fig:chis}),  
whereas the RPTA leads to a non-monotonic behavior
(see Fig. \ref{Fig:chis}). Establishing a 
microscopic connection between these two different approaches
is obviously a rather difficult and open issue. 

The LFFA is a semi-phenomenological approximation in which an analytical expression for the 
local field factor $G_{-}(q)$ is ``guessed'' with the constraints that the asymptotic behavior
should reproduce some  known results inferred from exact sum rules.
The unknown parameters of $G_{-}(q)$ are fixed from a fit to QMC data.\cite{davoudi:2001}
One may wonder whether one can extract some information about the possible shapes 
of $\chi_s(q)$ directly from the original QMC data. 
The QMC data shows a rather complicated structure with two extrema for $G_{-}(q)$ around $2k_F$
(see  Ref. \onlinecite{qmc} or Ref. \onlinecite{GV} (p. 244) ).
Though it might be tempting to relate the double-extremum structure obtained by the RPTA
to the QMC results, it turns out to be not possible to extract the behavior of $\chi_s(q)$
from available QMC data for $G_-(q)$. New QMC calculations directly computing $\chi_s(q)$ instead of $G_-(q)$
are thus highly desirable.\cite{note_slope}

Finally, we have seen that the low $q$ dependence of $\delta\chi_s(q)$ mimics the   
temperature dependence of $\delta\chi_s(T)$ which is
in agreement
with the experiment by Prus {\it et al.} at low $T$.\cite{reznikov:2003}
These experimental features  may provide another, though indirect, consistency check of the RPTA.

\section{Magnetic properties of the nuclear spins}\label{sec:magnetic}

We assume in this section that some nuclear spin ordering actually takes place at low enough 
temperature
and analyze how this ordering is destroyed when the temperature is raised.

\subsection{Mean field approximation}

Since the interaction between nuclear spins is of RKKY type, the interaction is
ferromagnetic at short distance $d\ll k_F^{-1}$ (the large $q$ behavior of $\chi_s(q)$ is only weakly 
modified by e-e interactions).
Furthermore, many mean field calculations performed for the 3D Kondo lattice at low electron
density (neglecting e-e interactions though) predict a ferromagnetic ordering.\cite{lacroix,sigrist:1997} Assuming 
a low temperature ferromagnetic ordering of the nuclear spins seems therefore 
a reasonable assumption. 

We first recall the mean field results for completeness.\cite{simon:2007}
The Weiss mean field theory gives a Curie temperature
\beq\label{eq:mf}
T_c^{MF}=-\frac{I(I+1)}{3k_B}\frac{A^2}{4n_s}\chi_L(q=0),
\eeq
where $I$ is the nuclear spin value.

In 2D 
this mean field theory yields  for $T_c^{MF}$ a  dependence
on the ratio $n_e/n_s$. 
For a  metal with about one conduction electron per nuclear spin, the ratio $n_e/n_s\sim 1$, and
we recover the result derived more than sixty years ago by Fr\"ohlich and Nabarro for a 3D bulk metal.\cite{FN}
For a 2D metal 
the Weiss mean field theory then gives $k_BT_c=I(I+1)A^2/12E_F$.  
For a 2D semiconductor, however, the smaller Fermi energy is compensated by
the smaller ratio $n_e/n_s \ll 1$. With typical values for GaAs heterostructures,
$I=3/2$, $A\sim 90~\mu eV$ and $a\sim$ 2\AA, \cite{coish:2004a} we estimate $T_c\sim 1~\mu K$, which is very low. 
(For such low $T_{c}$'s, ignoring nuclear dipole-dipole interactions from the start would 
not be legitimate.) 
However, this estimate is just based on the simplest mean field theory and, moreover, does not include the effect 
of e-e interactions. 
It still leads to a finite $T_c$ under which the nuclear spins
order ferromagnetically.

\subsection{Spin wave analysis around  ferromagnetic ground state}
\label{sec:fm}

We shall now go beyond the above mean field approximation and perform a spin wave analysis. 
The collective low-energy excitations in a ferromagnet
are then given by spin waves (magnon excitations).

\subsubsection{Magnetization and  Curie temperature of the nuclear spins}
>From standard spin waves analysis \cite{mermin},
the dispersion relation of the spin waves in the ferromagnet simply reads
\beq
 \omega_q=I(J_0-J_{q})=I \frac{A^2}{4}a^2(\chi_{s}({q})-\chi_{s}(0)),
\eeq
where $J_{\mathbf q}$ is the Fourier transform of $J_{\mathbf r}$ defined in Eq. (\ref{eq:J}).

At this stage, we already see that the stability of the ferromagnetic ground state 
demands that  $\delta\chi_s(q)=\chi_{s}({q})-\chi_{s}(0)>0$.
We can therefore  conclude that the second order calculation 
implies that the ferromagnetic ground state is always unstable.\cite{meaculpa} 
On the other hand, 
the renormalized perturbation theory approximation developed in Sec. \ref{sec:RPTA}
shows that it is necessary to go beyond lowest order perturbation theory.

When renormalization effects are not important in the RPTA, the lowest order perturbative 
results are recovered, and the ferromagnetic ground state seems unstable (though renormalization
effects in all channels must be carefully taken into account as described in Ref. \onlinecite{shekhter2:2006}).
When renormalization effects in the Cooper channel are important, 
we expect the two possible shapes denoted by (a) and (b) 
for the static spin susceptibility $\chi_s(q)$ at $T=0$
(see Fig. \ref{Fig:chis}). If the case (b) is favored, then there exists a value of $q$ 
at which $\omega_q<0$ signaling  an instability of the ferromagnetic ground state.
>From this perspective, cases (b) and (c) are similar.
Another ground state must then be assumed and a subsequent analysis is required.
This will be detailed in Sec. \ref{sec:helical}.

On the other hand, if the shape of the susceptibility denoted by (a) is favored, the ferromagnetic assumption
is  self-consistent. 
The RPTA predicts that there exists a temperature $T_0$ above which
$d\chi_s/dq<0$ at low $q$. 
This implies that there exists another temperature $T_1\leq T_0$
at which the minimum in $q_1$ touches the horizontal axis signaling an instability.
If the Curie temperature $T_c$ is larger than $T_1$, 
then there exists a temperature regime (typically for $T>T_1$) 
where the ferromagnetic ground state becomes unstable and a different 
ordering may be favored. This case will be analyzed in Sec. \ref{sec:helical}.
On the other hand, if  the Curie temperature $T_c$ is smaller than  $T_1$,
the ferromagnetic ground state is self-consistent below $T_c$ \cite{FootnoteT0}. Such a scenario
is in accordance 
with the one obtained from LFFA.
Let us therefore analyze this latter case.

The magnetization $m$ per site for a ferromagnet at finite $T$ is  defined by
\beq
m(T)=I-\frac{1}{N}\sum_{\mathbf q} n_q=I-\frac{1}{N}\sum_{\mathbf q} \frac{1}{e^{\beta\omega_q}-1},
\eeq
where $n_q$
is the magnon occupation number and the summation is over the first Brillouin zone of nuclear 
spins. In the continuum limit this becomess
\beq\label{eq:m}
m(T)=I- a^2\int \frac{d\mathbf q}{(2\pi)^2}\frac{1}{e^{\beta\omega_q}-1}.
\eeq
We define the Curie temperature  $T_{c}$ as the temperature
at which the magnetic order is destroyed by those spin waves. 
This procedure is equivalent to the
Tyablikov decoupling scheme. \cite{Tyablikov}
Another way of determining the Curie temperature  is to analyze at which
temperature $T_c$ the spin wave analysis breaks down.
The Curie temperature $T_c$ may be then defined by $m(T_c)=0$, 
which can be written as
\beq\label{eq:tc}
1=\frac{a^2}{I}\int \frac{d\mathbf q}{(2\pi)^2} \frac{1}{e^{ \omega_q/k_{B}T_{c}}-1}.
\eeq

For non-interacting electrons in 2D, $\chi_{s}(q)-\chi_{s}(0)=0$ for $q< 2k_F$,\cite{GV}
where $k_F$ is the Fermi wave vector. The spin wave analysis, therefore, gives $T_c=0$. 
This is in agreement with a recent conjecture extending the Mermin-Wagner theorem\cite {MW}
for RKKY interactions to  a non-interacting 2D system. \cite{bruno:2001}
For interacting electrons, however, the long range decay of the RKKY interactions can be altered
substantially and no conclusion can be  drawn from the Mermin Wagner theorem or
its extensions.

Let us now include electron-electron interactions (obtained either by the RPTA or LFFA). 
All approximations imply that the magnon dispersion is linear in $q$ at low $q$, 
{\it i.e.}
that $\omega_q\approx c q$, where 
\beq
c=I\frac{A^2}{4}a^2\left.\frac{\partial \chi_s(q)}{\partial q}\right|_{q\to 0},
\eeq
can be regarded as the spin wave velocity.\cite{note_cq} 
Such linear spin wave behavior is usually associated
with antiferromagnets while one would expect a quadratic dispersion for ferromagnetically ordered
states like considered here. 
This somewhat unexpected linear dispersion comes purely from  electron-electron interactions. 

The perturbative calculations or their extensions to include the Cooper pair instability
allows us to extract only the low $q$ asymptotic behavior of $\delta\chi_s(q)$. 
Monte Carlo results, however, seem to indicate that the local field factor $G_{-}(q)$ is almost linear in
$q$ up to $q \sim O(2k_F)$.\cite{GV} We will therefore assume that $\omega_q\approx c q$ for 
$q< q^*\sim O(k_F)$.

This implies that for $T<T^*$, where
\beq
T^*=c q^*/k_B,
\eeq
the integral determining $m$ in Eq. (\ref{eq:m})
is entirely dominated by
the linear dispersion behavior. Since fast modes corresponding to $q\gg q^*$ are exponentially
suppressed, we can easily compute it assuming $\omega_q$ linear in $q$ for the whole $q$ range
(extending the upper integration limit to infinity).
We obtain
\beq\label{eq:mt}
m(T)=I\left[1-(T/T_c)^2\right]~~{\rm for}~~T<T^*,
\eeq
where
\beq \label{eq:tcsw}
T_c^{}=\frac{2c}{k_B a}\sqrt{\frac{3 I }{\pi}}=\frac{A^2 I}{2k_B}\sqrt{ \frac{3I}{\pi n_s}}\left.
\frac{\partial \chi_s(q)}{\partial q}\right|_{q\to 0}
\eeq 
is the Curie temperature.
Note that with these estimates one has
\beq\label{eq:T*vTc}
\frac{T^*}{T_c}=\frac{a q^*}{2\sqrt{3I/\pi}}\ll 1~.
\eeq

Such a definition of $T_c$ has been obtained assuming that $\omega_q\approx cq$ for all $q$. 
This approximation
has two major aspects: First it
regularizes naturally the integral in Eq. (\ref{eq:tc}) in the UV limit, second only the low energy 
dependence of $\omega_q$ is taken into account, which is consistent with a spin wave approximation.

\subsubsection{Alternative UV regularization schemes }
In the previous section, we have assumed that $\omega_q\approx c q$ for all $q$.
On the other hand, one can assume we know explicitly $\omega_q$ for all $q$ in
the first Brillouin zone {\it i.e.} for
$q<\pi/a$ despite only the asymptotic limits of $\delta\chi_s(q)$ (and therefore of $\omega_q$)
are well established. At large $q$, we  expect electron-electron interactions to play a minor role
and the electron spin susceptibility to be well approximated by its non-interacting value,
which decreases as $1/q^2$ (see Eq. (\ref{eq:lindhard})).
This implies that $\delta\chi_s(q)\approx -\chi_s(0)=\chi_S$ at large $q$ and that the integral
in Eqs. (\ref{eq:m}) or (\ref{eq:tc}) are actually diverging when $a\to 0$.
If we adopt such a procedure, the integral  in Eq. (\ref{eq:tc}) 
is fully dominated by the short-distance modes,
{\it i.e.} by the UV cut-off (and therefore independent of any e-e interactions). 
Such a regularization scheme is not very satisfying and furthermore even inconsistent 
for a spin-wave approximation which relies on the long-ranged modes. 
Note that the $T_c$
we obtain with such procedure is similar (up to a prefactor of order unity) 
to the Curie temperature obtained within
the mean field theory in Eq. (\ref{eq:mf}).

Another regularization scheme consists in cutting off 
the integral in Eq. (\ref{eq:tc}) to $q<2 \zeta_F k_F$ with $\zeta_F$ a constant larger than $1$. 
This can be justified by 
integrating out 
fast modes  directly at the Hamiltonian level
in Eq. (\ref{eq:ueff}) since $\chi(q)$ decreases as $1/q^2$ for $q\gg k_F$.
Such a reasoning is equivalent in real space to a decimation procedure in which a square 
plaquette containing 
$(\zeta_F a k_F)^{-1}\times (\zeta_F a k_F)^{-1}$ nuclear spins is replaced by another plaquette
with a single average spin. Since at short distance, the RKKY interaction is mainly ferromagnetic
this is equivalent to a mean field procedure. The long distance interaction is not substantially 
modified. The main effect of this integration over fast modes is that the UV cut-off in 
Eq. (\ref{eq:tc}) is now of order 
$2 k_F$ instead of $\pi/a$. Although such a procedure does not allow for
an exact calculation of $T_c$ since $\omega_q$ is not known around $2k_F$, this considerably
boost the Curie temperature by orders of magnitudes compared to the previous regularization scheme
and  in the same range as the Curie temperature determined from Eq. (\ref{eq:tcsw}).

In the following, we will therefore use Eq. (\ref{eq:tcsw}) as our definition of $T_c$. This has 
the advantage of providing us with a simple closed formula. Furthermore, such a $T_c$ is consistent
with the long range approximation for spin waves.

\subsubsection{Numerical estimate of the Curie temperature}\label{sec:numerical}

Eq. (\ref{eq:tcsw}) gives us an estimate of the Curie temperature
as a function of the derivative of the electron spin susceptibility.
We have computed these quantities in Sec. \ref{sec:interaction} 
for various approximate schemes. 
We are therefore now ready to give estimates for the Curie temperature $T_{c}$.
Let us start with the RPTA. Assuming $\Gamma_c\sim O(1)$ in Eq. (\ref{eq:gamma_cutoff}),
we obtain a Curie temperature
$T_c^{}\sim 20~\mu K$ for  typical GaAs 2DEG parameters with $r_s\sim 1$.  
Setting $q^*\sim 2k_F$  in Eq. (\ref{eq:T*vTc}) and using the same parameters, one obtains
$T^*\sim T_c/50$.

At larger $r_s$,
$T_c$ is enhanced for two reasons: first $k_F^{-1}$ increases linearly with $r_s$,
and second,  the value of the spin susceptibility at $q=0$,
$\chi_S=|\chi_s(0)|$,
which is essentially the Pauli susceptibility at small $r_s$, increases linearly with $r_s$.\cite{GV}
An approximate  value of $\chi_S$ can be extracted from QMC calculations.\cite{GV} 
One obtains, for example, $T_c\sim 0.3~ mK$
for $r_s=5$ and $T_c\sim 0.7~mK$ for $r_s=8$. One may even obtain larger values of $T_c$
in the $mK$ range for larger $r_s$ since $T_c$ increases quadratically with $r_s$.
Furthermore, 
when $UN_e$ is no longer negligible compared to $1$, $T_c$ (resp. $T^*$) 
is even further enhanced by an additional  factor $1/(1-UN_e)^2$ (see Eq. (\ref{eq:deltachis})).
Close to the  ferromagnetic Stoner instability of the electron system, reached when $UN_e\sim 1$,
the Curie temperature $T_c$ for the nuclear system is strongly enhanced as could have been anticipated.

On the other hand, if we use the
local field factor approximations developed in Sec. \ref{sec:lffa}, 
we can determine $T_{c}$ by inserting $\chi_s(q)$ obtained from the LFFA 
into Eq. (\ref{eq:tcsw}):
\beq \label{eq:tcsw1}
T_{c}^{}=\frac{IA}{2k_B}\sqrt{ \frac{3I}{\pi }}\frac{A}{(\al-1)^2g_i {V}(a)},
\eeq 
where $g_i=g_0$ [Eq. (\ref{eq:g0})] or $g_i=g_1$ [Eq. (\ref{eq:g1})].  
The energy scale $(\al-1)^2g_i {V}(a)$ can be interpreted as a renormalized screened potential
due to collective interaction effects that are incorporated in the LFFA.

If we use $g_i=g_0$, the LFFA  gives an exponential enhancement of $T_{c}$ 
with increasing the interaction parameter $r_s$. Yet, as we have discussed in Sec. \ref{sec:lffa}, 
this form for the
local field factor $G_-$ cannot really be trusted at large $r_s>1$ when compared to QMC data.
If we instead use the MLFFA, which seems to be in better agreement with QMC data, then $g_i=g_1$ and
we obtain $T_c\sim 0.4~mK$ for $r_s\sim 5$ and $T_c\sim 1~mK$ for $r_s\sim 8$.
These values are consistent with the ones found using renormalized perturbation theory.

Note that the ratio ${A}/(\al-1)^2g_0 {V}(a)$ can be regarded as the small parameter of our theory.
For some large value of $r_s$, the dimensionless parameter ${A}/(\al-1)^2g {V}(a)$
may approach unity and the truncation of the Schrieffer-Wolff 
transformation at lowest order becomes unjustified.

\subsection{Self-consistent calculation of the Curie temperature}

If we assume a ferromagnetic ordering of the nuclear spins, this generates some
rather large nuclear Overhauser magnetic field $B_{\rm eff}(m)\sim O(1~\mathrm{T})$ in a  GaAs 2DEG for $m\lesssim I$
that we have neglected before. 
Therefore, the spin degeneracy of the
conduction electrons will be lifted by the Zeeman energy $g\mu_B B_{\rm eff}$
which should have some effects on the determination of $T_c$, too. A self-consistent procedure
including feedback effects is therefore required. This is the purpose of this section.

The electron spin susceptibility is no longer isotropic. In the following we will assume
that it is still diagonal but has now a longitudinal component $\chi_{s}^{z}$ (in the direction of the magnetic field produced by the nuclear spins) 
and a transverse component $\chi_{s}^{\perp}$. Taking into account such an anisotropy, our SW 
transformation 
is still valid and we obtain
an effective spin anisotropic 2D long-ranged Hamiltonian, which replaces Eq. (\ref{eq:hreal}):
\beq
H_{\rm eff}=-\frac{1}{2}\sum\limits_{\mathbf r,\mathbf r'} 
\left[J_{|\mathbf r-\mathbf r'|}^z I_{\mathbf r}^zI^z_{\mathbf r'}
+\frac{J_{|\mathbf r-\mathbf r'|}^\perp}{2} (I_{\mathbf r}^+I^-_{\mathbf r'}+I_{\mathbf r}^-I^+_{\mathbf r'})\right],
\eeq
where, in $q$ space, $J_q^{z/\perp}=-(A^2/4n_s)\chi^{z/\perp}(q)$.
The dispersion relation for magnons becomes now
\beq
\omega_q=I(J^z_0-J^\perp_q),
\eeq
and therefore generically acquires a gap for $B_{\rm eff}\ne 0$. 
Let us denote by $\Delta(B_{\rm eff})=\Delta(m)$ such a gap. 
The expansion of the magnon dispersion  for small $q$ leads to
\beq
\omega_q\simeq I\Delta+ I q \left.\frac{\partial J^\perp_q}{\p q}\right|_{q=0}=I\Delta+ c' q,
\eeq
where we have assumed a linear in $q$ behavior for $\chi_s^\perp(q)$ and defined 
$c'=I \frac{\partial J^\perp_q}{\p q}(q=0)$. We assume ferromagnetic order and so
$c'>0$.
In the following we develop a simple approach of the Landau type.
Eq. (\ref{eq:tc}) which determines the Curie temperature is still valid and now reads:
\beq
1=\frac{a^2}{I}\int \frac{d\mathbf q}{(2\pi)^2} \frac{1}{e^{\beta I\Delta+\beta c'|\mathbf q| }-1 },
\eeq
where $\beta=1/k_B T$.
As before, we assume the linear dispersion to be valid for all $q$, which naturally regularizes the
integral in the UV limit. The latter integral equation can be rewritten in a more compact form as
\beq\label{eq:tcgap}
1=\frac{a^2}{2\pi I}\frac{1}{(\beta c')^2}{\rm Li}_2(e^{-\beta I \Delta(m(T))}),
\eeq
where ${\rm Li_2}$ is the dilogarithm function.
Let us first focus on the magnetization $m$.
>From Eq. (\ref{eq:m}) we have
\beq \label{eq:m_Li}
m(T)=I-\frac{a^2}{2\pi}\frac{1}{(\beta c')^2} {\rm Li}_2(e^{-\beta I \Delta(m(T)) }).
\eeq
When $T\to T_c$, $m\to 0$ and hence $\Delta(m)\to 0$. Let us denote by $T_c^0$ the Curie temperature
obtained from Eq. (\ref{eq:tc}) by neglecting $B_{\rm eff}$. We assume that the temperature
is below but close to $T_c$. We can therefore expand the gap as $\Delta=\Delta'm+\cdots$
with $\Delta'=\frac{d\Delta}{dm}(m=0)$. 
The dilogarithm around $\Delta=0$ then expands to:
\bea
{\rm Li}_2(e^{-\beta I \Delta' m})&=&\frac{\pi^2}{6}+g(\beta I \Delta' m)
-\frac{(\beta I \Delta')^2}{4}m^2\nn\\
&&+\frac{(\beta I \Delta' )^3}{72}m^3+\cdots~,
\eea
where $g(x)=x(\ln x-1)$.
Introducing the dimensionless parameters $t=T/T_c-1$, $t_0=T/T_c^0-1$, $b=I\Delta'/k_B T_c$,
the self-consistent equation (\ref{eq:m_Li}) for $m(T)$ becomes
\bea
m(t)&=&-2t_0+\frac{6}{\pi^2} b m \left(1-\ln (b m)\right)+\frac{3b^2}{2\pi^2}m^2\nn\\
&&-\frac{b^3}{12\pi^2}m^3+\cdots~.
\eea
This equation can be easily solved numerically. Note that by integrating this equation
with respect to $m$, one obtains the Landau functional.

\begin{figure}[t]
\hskip -1.cm \psfig{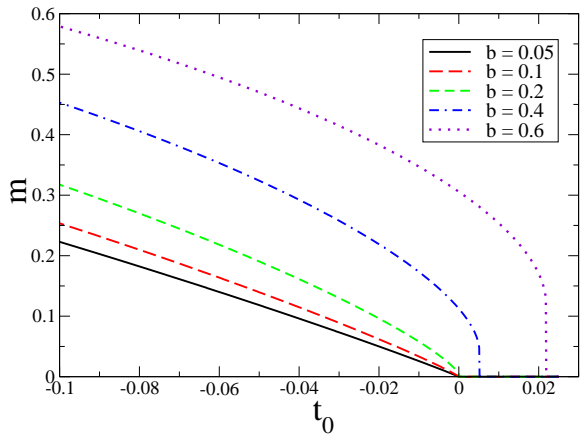}
\caption{ Magnetization as a function of $t_0=(T/T_c^0)-1$ for various values
of the dimensionless parameter $b$. For small values of $b$, the behavior described by Eq. (\ref{eq:mt}) is recovered.}
\label{Fig:mt0}
\end{figure}
In Fig. \ref{Fig:mt0}, we plotted the magnetization as a function of $t_0$ for various values of
the dimensionless parameters $b$. Two interesting features can be noticed: First, $T_c$ increases
compared to $T_c^0$ by increasing $b$. Second, the magnetization rises sharply as $T$ is lowered
through $T_c$, which is reminiscent of a first order phase transition for a system in the
thermodynamic limit. This effect becomes pronounced at values of $b \sim 0.15$.

Let us now estimate the value of $b$ in our system. 
To this end, we need to estimate $\Delta'$ which can be regarded as the susceptibility of 
the susceptibility. Since the limits $T\to 0$ and $B_{\rm eff} \to 0$ for the susceptibility 
$\chi_s(T,B_{\rm eff})$
do not commute,\cite{shekhter2:2006} a little care is required.
When the nuclear spins in the  GaAs heterostructure are polarized, they generate a rather 
large effective magnetic field $B_{eff}$ which gives an electronic
Zeeman energy scale $\Delta(m=1)$ of the order $I A\sim O(1~\mathrm{K})$. This scale 
is much larger than the typical Curie temperature we estimated before. 
In the ferromagnetic phase, we are therefore
in a  regime in which $\Delta(m=1)\gg k_BT$, where $T$ is the temperature.

In this regime, one can use Ref. \onlinecite{shekhter2:2006} to estimate
$\Delta'\sim A^3/E_F^2 (ak_F)^2 \Gamma_c^2$, where $\Gamma_c$ is the renormalized scattering amplitude
defined in (\ref{eq:gamma0}). Using Eq. (\ref{eq:tc}) for the Curie temperature $T_c$,
we can estimate $b\sim (ak_F) (A/E_F)\ll 1$. Our problem thus corresponds to a regime in which 
feedback
effects are negligible, which justifies the previous isotropic approximation for the electron spin
susceptibility.

\subsection{Spin wave analysis around a helical ground state}
\label{sec:helical}

We have assumed so far that the ground state is ferromagnetic.
This assumption obviously depends on the sign of $\delta \chi_s(q)$.
We have seen that the RPTA predicts two possible scenarios: Either $\delta \chi_s(q)<0$ in some 
range of $q$ (case (b) in Fig. \ref{Fig:chis}) or $\delta\chi_s(q)$ becomes negative 
above some temperature 
$T_1$. A positive sign, on the other hand, is  always obtained from LFFA.
In contrast, second order perturbation theory gives a negative sign
for $\delta \chi_s(q)$, independent of the magnitude of $q$. 
If we assume that there is a temperature range where
$\delta\chi_s(q)<0$, this clearly implies that the assumption of ferromagnetic 
order is invalid since the magnon spectrum has $\omega_q<0$. 
This corresponds to the situations (b) or (c) in Fig.~\ref{Fig:chis}, 
which may depend on the Fermi liquid parameters or arise when $T$ exceeds
some temperature $T_1$.

\subsubsection{A new ground state}

A different ground state for the nuclear spins thus has to be assumed.
If the minimum of $\delta \chi_s(q)$ is reached at some finite wave vector 
$q_1 \sim k_F$, a reasonable ground state of such a system
could be an incommensurable helical state, or a spiral state,
with a wavelength of the helix of the order of 
$\lambda_1 = 2\pi/q_1$.

We stress that, {\it on the scale of the lattice spacing}, the RKKY interaction remains ferromagnetic,
and the neighboring nuclear spins are still ferromagnetically aligned. On the scale of the 
Fermi wavelength, however, opposite spin alignment is favored.
A gain in energy may be obtained by a slow variation of the nuclear spin alignment, 
i.e. by helical order. If such a nuclear spin arrangement is reached for some temperature range, 
it implies that
the total magnetization vanishes over distances much longer than $\lambda_F$.


We assume that the ground state of the nuclear spins 
is described by a local magnetization
$\mathbf{m}_i = \langle \mathbf{I}_i \rangle$
that is confined to the 
spin $(x,y)$ plane and that rotates in spin space
around the  $z$-direction.
A full rotation is described by the 
wavevector $\mathbf{q}_1$. 

These $\mathbf{m}_i$ play the role of the order parameter in the system.
In order to compare with a ferromagnet, we can characterize the order
by locally rotating the $\mathbf{m}_i$ so that they all map onto 
the same $\mathbf{m}_0$:
\begin{equation} \label{eq:rot_m}
	\mathbf{m}_i = R_i \mathbf{m}_0 
	= \left(\begin{matrix} 
	\cos(\mathbf{r}_i \cdot \mathbf{q}_1) & \sin(\mathbf{r}_i \cdot \mathbf{q}_1) & 0 
	\\
	- \sin(\mathbf{r}_i \cdot \mathbf{q}_1) & \cos(\mathbf{r}_i \cdot \mathbf{q}_1) & 0 
	\\
	0 & 0 & 1
	\end{matrix} 
	\right) \mathbf{m}_0,
\end{equation} 
where  $\mathbf{m}_0=\mathbf{m}(\mathbf r_i=0)$ is the magnetization at site $\mathbf{r}_i=0$.
Note that  $\mathbf{m}_0$ has 
the same interpretation as the average magnetization of a ferromagnet.

The helical
ground state breaks the translational as well as the rotational symmetry 
of the Hamiltonian in Eq. (\ref{eq:ueff}). In particular, the ground state is degenerate with respect to
a change of the direction of $\mathbf{q}_1$ in the 2DEG plane.
With the choice of a specific direction $\mathbf{q}_1$, this symmetry is broken. 
There exists, therefore, a Goldstone mode that tends to restore this
rotational symmetry by inducing changes on the direction of $\mathbf{q}_1$.
Obviously, such a mode would destroy the assumed order immediately.
Realistically, however, the specific helical state can be pinned by the 
system, such as by a disorder lattice configuration or by some Dzyaloshinskii-Moryia interactions,
and the Goldstone
mode becomes massive.\cite{belitz:2006} In the following we will assume that this is the 
case. 
Concretely this means that excitations with small $\mathbf{q}$ 
in the perpendicular direction of $\mathbf{q}_1$ are strongly suppressed.

\subsubsection{Spin wave analysis}
Our starting point is the effective spin Hamiltonian of Eq.~(\ref{eq:hreal})
with an exchange interaction matrix $J^{\alpha\beta}_{ij} = J^{\alpha\beta}(|\mathbf{r}_i-\mathbf{r}_j|)$
defined in Eq. (\ref{eq:J})
that can be off-diagonal in general.
Since $J^{\alpha\beta}_{ij}$ is proportional to the spin susceptibility,
as can be seen from Eq. (\ref{eq:susc}),
we can assume, however, that $J^{xx}_{ij} = J^{yy}_{ij}$ and that 
$J^{\alpha\beta}_{ij} = -J^{\beta\alpha}_{ij}$ for $\alpha \neq \beta$. 
These assumptions are related to the conservation of the total spin of the electron system.\cite{note_cons}

A convenient way to perform the spin wave analysis is to first perform a local rotation 
of each nuclear spins $I_j$ as in Eq. (\ref{eq:rot_m}) such that they all 
become parallel to each other like in a ferromagnet.
We thus define a local (right-handed) set of axes described by the unit vectors
$\mathbf{e}_i^1, \mathbf{e}_i^2, \mathbf{e}_i^3$, 
with $\mathbf{e}_i^1$ being parallel to $\mathbf{m}_i$.\cite{schuetz}
In principle, $\mathbf{e}_i^2$ and $\mathbf{e}_i^3$
can be chosen arbitrarily. It is convenient though to choose $\mathbf{e}_i^3$ parallel to
the spin rotation axis $z$.
Then we can write
$\mathbf{I}_i = \tilde{I}_i^1 \mathbf{e}_i^1 + \tilde{I}_i^2 \mathbf{e}_i^2 + \tilde{I}_i^3 \mathbf{e}_i^3$.
These new components $\tilde{I}_i^\alpha$ are connected to the original components $I^{\alpha}_i$ through
the matrices $R_i$ as
$(\tilde{I}_i^1,\tilde{I}_i^2,\tilde{I}_i^2)^T = R_i^\dagger (I_i^x,I_i^y,I_i^z)^T$,
where $T$ denotes here the transposition.
Let $\mathbf{J}_{ij}$  be the $3 \times 3$ matrix associated with $J^{\alpha\beta}_{ij}$,
then the Hamiltonian for the nuclear spins can be written as
\begin{equation} \label{eq:heli_Heff}
	H_{\rm eff} = - \frac{1}{2}\sum_{ij} (\tilde{I}_i^1,\tilde{I}_i^2,\tilde{I}_i^3)
	R_i \mathbf{J}_{ij} R_j^\dagger 
	\left(\begin{matrix} 
	\tilde{I}_j^1\\ \tilde{I}_j^2 \\\tilde{I}_j^3\end{matrix}\right).
\end{equation} 
In this new basis, the spin-wave analysis is analogous to the ferromagnetic case and  
rather standard (see e.g. Ref. \onlinecite{maleyev}).

The ground state energy of the helimagnet then becomes
\beq\label{eq:e0hel}
E_0= - \frac{I^2 N}{2} J^{xx}_{q_1}.
\eeq
Let us compare Eq. (\ref{eq:e0hel}) to the ground state energy  of a 
ferromagnet. If all spins are aligned
along the $x$ direction the ground state energy
of the ferromagnet is
\begin{equation} \label{eq:Efm}
	E_0^{\rm FM} = -\frac{I^2}{2} \sum_{ij} J_{ij}^{xx}
	= -\frac{I^2 N}{2} J^{xx}_{q=0}.
\end{equation}
The energy of the helical state is thus lower than that
of the ferromagnet if $J^{xx}_{q_1} > J^{xx}_{q=0}$.
Furthermore, the helical state has the lowest energy 
at the wave vector $\mathbf q_1$, where $J^{xx}_{q} = J^{yy}_{q}$ has its 
maximum.

The low energy excitations above the ground state can also be obtained in a straightforward
manner from Eq.  (\ref{eq:heli_Heff}). We find the following two branches of the spin wave spectrum,
\begin{eqnarray}
	\omega_{q}^{(1)} 
	&=& \frac{I}{2} (J^{xx}_{q_1} - J^{xx}_{|\mathbf{q}_1+\mathbf{q}|}),
\label{eq:omega_1}
\\
	\omega_{q}^{(2)} 
	&=& \frac{I}{2} (J^{xx}_{{q}_1} - J^{zz}_{q}).
\end{eqnarray}
Clearly, $\omega_\mathbf{q}^{(1)} = 0$ at $\mathbf{q}=0$. On the other hand, we 
see that the helical ground state can only be stable if $\omega_{\mathbf{q}}^{(2)} \ge 0$,
which means that $J^{zz}_{|\mathbf{q}|}$ must not exceed 
$J^{xx}_{q_1}$ in the vicinity of $\mathbf{q} \approx 0$.
If $J^{zz}_q = J^{xx}_q$ this is indeed the case. The 
second branch, $\omega_\mathbf{q}^{(2)}$, then has a gap.

\subsubsection{Effect of gapless modes}

We see
that there is a gapless mode $\omega_{\mathbf q}$ in the system, given by 
Eq.~(\ref{eq:omega_1}).
If $\mathbf{q}$ is such that $|\mathbf{q}_1+\mathbf{q}| = q_1$, then 
$\omega_\mathbf{q}$ remains strictly zero. This is the aforementioned
Goldstone mode. 
Such fluctuations, therefore, are assumed to acquire a mass, associated with an energy scale $\Delta_G$.
In the longitudinal direction, however, where $\mathbf{q}$ is 
parallel to $\mathbf{q}_1$, $\omega_{\mathbf{q}}$ grows
 for $|\mathbf{q}_1+\mathbf{q}|>q_1$ and $|\mathbf{q}_1+\mathbf{q}|<q_1$
proportionally to the increase of $\chi_s(|\mathbf{q}_1+\mathbf{q}|)$ with respect to 
its minimum.

The minima of the spin wave spectrum are pushed to 
finite momenta $q$, notably to the minimum of the spin susceptibility
$\chi_s(q)$ at finite $q=q_1$. Provided $\chi_s(q)$ remains
analytic around the minimum at $q_1$, it no longer grows linearly but, in general, as 
$(q-q_1)^2$. The system, therefore, can no longer benefit from a linear $\chi_s(q)$ 
to stabilize the ferromagnetic order and different, nonuniversal, 
energy and length scales will affect the Curie temperature $T_c$ (see below).

We have seen above that the quantity $m_0 = |\mathbf{m}_0|$ can be used as the order
parameter of the helical state. Every spin wave reduces $m_0$,
and we can thus use again Eq.~(\ref{eq:m}) in order to express 
how the average local magnetization is reduced by the spin excitations,
\begin{equation} \label{eq:m_q0}
	m_{0}
	= I - a^2 \int \frac{d\mathbf{q}}{(2\pi)^2} \frac{1}{e^{\beta \omega_{q}}-1}.
\end{equation}
For $q \to 0$, the integrand becomes singular as $1/q^2$.
\cite{heli_nonanalytic}
The singularity
cannot be compensated by the factor $q$ of the spherical integration measure 
$q \, dq$ as in the previously discussed ferromagnetic case.
This is the same situation that is met when
calculating spin wave excitations for systems without long-ranged interactions.
There the singularity is directly linked to the Mermin-Wagner theorem. \cite{MW} 
In the present case, however, the small $q$ values are cut off at some
finite inverse length scale $\pi /L$, associated with the energy $\Delta$,
the infrared cut-off frequency introduced in Sec. \ref{sec:interaction}.

The scale $L$ lifts the singularity in the integral (\ref{eq:m_q0}) 
by cutting off the momentum at $q \approx \pi/L$, 
and so by effectively opening a gap for the excitations. 
The spin system can maintain a quasi-order over the length $L$. 
An additional reduction of the singularity also arises from the fact that
 the system is not truly 2D but a layer with a finite width $w_z$, containing
 about 50-100 planes of nuclear spins (see also App. A).
This length scale, however, must be compared with the typically much longer
wavelengths of the spin waves, and thus can only account for a partial
regularization.
The singularity is dominated by the minimal curvature of $\omega_\mathbf{q}$
in all directions of $\mathbf{q}$. In the present case there are two main
directions, the longitudinal one parallel to $\mathbf{q}_1$, where
$\omega_\mathbf{q} = C q^2$, with 
$C = -\partial^2 J_{ q}/\partial q^2|_{q= q_1}$, 
and the transverse one perpendicular to $\mathbf{q}_1$, where the curvature
is imposed by $\Delta_G$. For the stability of the ground state we must assume
that the pinning strength $\Delta_G$ is large compared to the energies 
imposed by the $J^{\alpha\beta}_q$. The singularity in Eq. (\ref{eq:m_q0}) 
is thus dominated by $1/\beta C q^2$

Let us assume  $L \sim 10\mu\mbox{m}$. 
Cutting off the upper integration limit by $k_F$, the singular part of 
Eq. (\ref{eq:m_q0}) becomes
\begin{equation} \label{eq:heli_cutoff}
	\int_{\pi/L}^{k_F} \frac{dq}{C\beta q} 
	=
	\frac{1}{C\beta} \ln(k_F L / \pi).
\end{equation}
The logarithm yields a factor exceeding one. The helical order cannot be 
stable if the expression in Eq.~(\ref{eq:heli_cutoff}) becomes larger than one. 
This allows us to define a temperature
\begin{equation} \label{eq:TG}
	k_B T^{G} = \frac{C}{I a^2 \ln(k_F L / \pi)},
\end{equation}
above which the gapless mode definitely destroys the helical order.

We see that much of the stability depends on the value of 
$C$,
which means, on the curvature of $\chi_s(q)$ around its minimum at $q_1$.
We can very roughly estimate $C/a^2 \sim J_{q_1} / (a k_F)^2$, and see
that this temperature $T^G$ can actually be quite high.
Let us now write the integral in Eq.~(\ref{eq:m_q0}) in the form
\begin{multline} \label{eq:log_out}
	\int \frac{d\mathbf{q}}{(2\pi)^2} \frac{1}{e^{\beta \omega_\mathbf{q}}-1}	
	=
	\frac{\ln(k_F L/\pi)}{\beta C}
	\\
	+
	\int_{|\mathbf{q}|>k_F} \frac{d\mathbf{q}}{(2\pi)^2} \frac{1}{e^{\beta \omega_\mathbf{q}}-1} .	
\end{multline}
Let us now further introduce a temperature $T^{**}$, similar to the temperature $T^*$
for the ferromagnetic case, below which the integral determining $m_0$ is
entirely dominated by the quadratic dispersion behavior. 
This means we assume that $\omega_\mathbf{q} \approx C q^2$ (in the direction
parallel to $\mathbf{q}_1$) up to a $q^{**} < q_1, k_F$. We can set
\begin{equation}
	T^{**} = C (q^{**})^2 / k_B. 
\end{equation}
For $T_1 < T < T^{**}$, therefore, Eq. (\ref{eq:log_out})
is controlled entirely by the first logarithmic part. 
We thus obtain
\begin{equation}
	m_0 \approx I - a^2	\frac{\ln(k_F L/\pi)}{C} k_B T
	= I (1- T/T^G).
\end{equation}
valid for $T<T^{**}$ .

The temperature $T^G$ can be seen as a generalization of $T_c$ for the
helical case. It differs from the ferromagnetic case through its
dependence on external, non-universal cut-off scales. 
This loss of universality is an essential difference compared to the
 previously studied  ferromagnetic order.
This also indicates that
the helical order is much more fragile with respect to external conditions than
the ferromagnetic order. 

At higher temperatures the formation of
defects and magnetic domains will further tend to destabilize the order
as well. It is therefore possible that the helical order is
destroyed well below $T^G$.

\section{Conclusion and discussion}\label{sec:conclusion}

{\em Summary} In this paper we have examined the interplay between an interacting electron
liquid in 2D with magnetic order in a lattice of nuclear spins. 
We have  based our investigation on a Kondo lattice model, in which 
the electrons couple weakly  to the nuclear spins through the hyperfine interactions.
In this way, an effective coupling of the RKKY type is induced between
the nuclear spins which, as obtained through a Schrieffer-Wolff transformation,
is expressed in terms of the static electron spin susceptibility $\chi_s(q)$.

Electron-electron interactions in 2D can substantially modify the shape of 
$\chi_s(q)$ and therefore profoundly affect the magnetic properties of the nuclear spin system.
A magnetic order can arise because the conditions for the Mermin-Wagner theorem
are not met due to the long-range character of $\chi_s(r)$. 
Much depends thus on the precise shape of $\chi_s(q)$. Based on a renormalized
perturbation theory we argued that for short-ranged interactions, $\chi_s(q)$ 
should have the forms sketched
in Fig.~\ref{Fig:chis}. When renormalization of the scattering amplitudes is important, 
$\chi_s(q)$ has two extrema at values $q_0,q_1 \sim k_F$,
which lead to the two generic situations labeled by (a) and (b) in Fig. \ref{Fig:chis}. 
If renormalization effects are unimportant, this leads to the form (c) as sketched in  Fig.~\ref{Fig:chis}. 
The distinction between (a), (b), and (c) is non-universal and, presumably, depends
on the sample disorder, the interactions, and the temperature.

For long-ranged Coulomb interactions, on the other hand, a calculation 
based on a local field factor approximation produces a monotonic increase 
of $\chi_s(q)$ (not sketched in  Fig.~\ref{Fig:chis}). 

Such a monotonic increase or the case (a) stabilize a nuclear ferromagnet.
In the case (b) the ferromagnetic order has an instability
at the wave vector $q_1 \sim k_F$, which corresponds to the absolute minimum
of $\chi_s(q)$, but  ferromagnetic coupling is maintained at short 
distances (as compared to the Fermi wavelength).  Similar behavior emerges also for case (c).
We argued that the nuclear ground state then has a quasi-order
which is (nearly) ferromagnetic on short distances but rotates the local magnetization
on a scale of $1/q_1$, thus providing a helical order. 

{\em Experimental implications} 
The transition temperature $T_c$ describes the temperature above which the
magnetic order breaks down. While the mean field approximation predicts a very low
transition temperature of the order of $\sim 1 \mu\mbox{K}$, we have
seen that  electron-electron interactions can considerably 
increase the value of $T_c$. From our various approximation schemes
we obtain a $T_c$ in the $mK$ range for both short-ranged and long-ranged interactions
for $r_s=5-10$ (see Sec. \ref{sec:numerical}).

If temperature is decreased below $T_c$, the nuclear spins order and generate an
effective magnetic field $B_{\rm eff}$, which can be very large in GaAs 2DEGs,
this being in contrast to Si MOSFETs, which have a much smaller Overhauser field.
This internal magnetic field has important consequences for the thermodynamic behavior of the electron
spin susceptibility, which can be derived from the non-analytic dependence of
$\chi_s(T,B_{\rm eff})$ on temperature and magnetic field following 
Ref. \onlinecite{shekhter2:2006} (the limits $T\to 0$ and $B_{\rm eff}\to 0$ do not commute).
In a Si MOSFET 2DEG, 
we expect $g\mu_B B_{\rm eff}\ll k_BT_c$ and therefore the calculated and observed\cite{reznikov:2003} 
linear in $T$ behavior of $\chi_s(T)$ is also valid  below $T_c$.
However, in GaAs 2DEGs one has clearly  
$g\mu_B B_{\rm eff}\gg k_BT_c$, and one may therefore expect $(\chi_s(T)-\chi_s(0))\propto T$ 
above $T_c$ and
$(\chi_s(T)-\chi_s(0))\propto m\propto [1-(T/T_c)^2]$ below $T_c$
in contrast to Si MOSFETs. This implies an upturn of $\chi_s(T)$ around $T=T_c$ in a GaAs 2DEG.

{\em Open questions}
There remain many open questions. Mainly, a detailed study of 
$\chi_s(q)$ at values $q \sim 2k_F$ could clarify  the scenarios
qualitatively sketched in Fig.~\ref{Fig:chis}. 
Monte Carlo simulations and experiments may provide further insight here. 
It would be desirable to establish a  general magnetic phase diagram for the nuclear spins  
as a function of $r_s$ and $T$. Possible new phases such as a nuclear spin glass phase
are likely due to the complexity and richness of the problem. 
Finally, disorder may play an important role by providing a further cut-off length,
and the interplay with electron-electron interactions requires a separate investigation.

\begin{acknowledgments}
We would like to acknowledge fruitful discussions with A. Millis, 
G. Vignale,
and useful correspondence with A. Finkel'stein.
This work is supported by the Swiss NSF, NCCR Nanoscience, and JST ICORP.
\end{acknowledgments}

\appendix

\section{Derivation of the effective magnetic low energy Hamiltonian}\label{sec:SW}

\subsection{Effective Hamiltonian}

We start from the general Kondo lattice Hamiltonian in Eq. (\ref{eq:kl}) with $H_{dd}=0$.
Since $A$ is a small energy scale in our case, we can perform a Schrieffer-Wolff (SW) transformation
in order to eliminate terms linear in $A$, followed by ``integrating out'' the electronic
degrees of freedom.
Keeping the lowest order terms in $A^2$ of the SW transformation,
we are left with an effective interaction Hamiltonian $H_{SW}$ that reads 

\beq 
H_{SW}=H_0-\frac{1}{2}[S,[S,H_0]].
\eeq
$S$ is defined by $H_n+[S,H_0]=0$, which is solved as $S=L_0^{-1}H_n$,
where $L_0$ is the Liouvillian superoperator.

Let us define \beq
U=\frac{1}{2}[S,[S,H_0]],\eeq
which can be rewritten as $U=\frac{1}{2}[L_0^{-1}H_n,H_n]$.
Using an integral representation for  $L_0^{-1}=-i\int\limits_0^\infty dt e^{i(L_0+i\eta)t}$, 
one obtains
\beq 
U=-\frac{i}{2}\int_0^{\infty}dt e^{-\eta t} [H_n(t),H_n],
\eeq
where $H_n(t)=e^{iL_0t}H_n=e^{iH_0t}H_ne^{-iH_0t}$,
and $\eta \to 0^+$ ensures the convergence of the time integration.
Using the definition of $H_n$ given in Eq. (\ref{eq:kl1}),
we are left in $q$-space with
\bea \label{eq:U_unaveraged}
 U&=&-\frac{iA^2}{8N_l^2}\sum\limits_{\mathbf q,\mathbf q'}\int_0^{\infty}dt 
 e^{-\eta t}[\mathbf I_{\mathbf q}\cdot\mathbf S_{\mathbf q}(t),
\mathbf I_{\mathbf q'}\cdot\mathbf S_{\mathbf q'}]\nn\\
&=&-\frac{iA^2}{8N_l^2}\sum\limits_{\mathbf q,\mathbf q'}\int_0^{\infty}dt e^{-\eta t}
\left\{I_{\mathbf q}^\alpha I_{\mathbf q'}^\beta[S_{\mathbf q}^\alpha(t),S_{\mathbf q'}^\beta]\right.\\
&&\hskip 3cm +\left. [I_{\mathbf q}^\alpha, I_{\mathbf q'}^\beta] S_{\mathbf q'}^\beta S_{\mathbf q}^\alpha(t)\right\}.\nn
\eea
Summation over Greek indices is implied.
The first commutator enters the definition of the susceptibility in 
Eq. (\ref{eq:susc}) below.
The second commutator can be straightforwardly computed by going back to real space.
We obtain $[I_{\mathbf q}^\al,I_{q'}^\be]=i \varepsilon^{\al\be\ga}I^\ga_{\mathbf q+\mathbf q'}$ where
$\varepsilon^{\al\be\ga}$ is the fully antisymmetric tensor.

We next take the equilibrium expectation value over electronic
degrees of freedom only, denoted by $\la\dots\ra$.  
Furthermore, we assume translational invariance in the 2DEG,
which implies $\la O_{\mathbf q}O_{\mathbf q'}\ra=N\la O_{\mathbf q}O_{-\mathbf q}\ra\delta_{\mathbf q+\mathbf q',0}$, with $N$ being the number of  sites in a 2D lattice.
Since $H_0$ has time-reversal symmetry the term proportional to 
$\epsilon^{\alpha\beta\gamma} I^\gamma$  in Eq. (\ref{eq:U_unaveraged}) 
drops out.
Together with the reduction to a 2D problem (discussed 
below), this allows us to bring $\la U\ra$ to a much simpler form,
\beq \label{eq:ueff_appendix}
\la U\ra=\frac{A^2}{8 N}a^2\sum\limits_{\mathbf q} I_{\mathbf q}^\al~
\chi_{\al \be}( q) ~I^{\be} _{-\mathbf q}~,
\eeq
where $n_s=a^{-2}$ is the 2D nuclear spin density. The quantity
\beq \label{eq:susc}
\chi_{\al\be}(q,\omega)=-\frac{i}{Na^2}\int_0^\infty dt~ e^{-i \omega t-\eta t}
\la[ S_{\mathbf q}^\al(t),S_{-\mathbf q}^\beta]\ra
\eeq
is the  2D electron spin susceptibility which includes electron-electron interactions.

\subsection{Reduction to 2D}

As already mentioned, the electron gas is not strictly 2D but has a finite 
thickness, typically of the order of $w_z\sim 5~{\rm nm}$ and 
therefore contains several layers of 2D nuclear spin planes. 
However, using that the electron wave function is confined in the third dimension ($z$-direction), 
we can reduce  the original  3D system to an effective 2D one.
Let us suppose a lattice site is labeled by 
$\mathbf r_j=(\mathbf r_{j//},z_j)$, where $(\mathbf r_{j//}$ is the planar coordinate and $z_j$  
the position in the perpendicular direction. We consider $N_z\sim 50$ layers
such that $N_l=N\times N_z$.

If we assume that $w_z$ is sufficiently small  so that the electrons are confined
in a single mode $\phi(z)$ in the $z$-direction, all nuclear
spins in a column along the $z$-direction  couple to the same electron wave function. 
Fluctuations of the nuclear spins along this direction are expected to be weak. In a 
mean-field-like description, we may thus replace the column by
a single 2D nuclear spin as follows. Let us separate out the 
perpendicular mode from the electron spin operator as 
\beq
\mathbf{S}_i(\mathbf{r}_{i//}, z_i) \longrightarrow \frac{|\phi(z_i)|^2}{w_z} \mathbf{S}_i(\mathbf{r}_{i//}),
\eeq 
with
$\mathbf{S}_i(\mathbf{r}_{i//})$ being a  2D electron spin operator.
The mode $\phi(z_i)$ can then be used as an envelope function for the 
nuclear spins 
\beq
\mathbf{I}_i(\mathbf{r}_{i//}) = \frac{1}{w_z}\int dz_i |\phi(z_i)|^2 \mathbf{I}_i(\mathbf{r}_{i//},z_i).\eeq
Since $w_z$ is determined by $w_z = \int dz |\phi(z)|^2$, these new operators 
$\mathbf{I}_i(\mathbf{r}_{i//})$ satisfy the standard spin commutation relations.
The remaining Hamiltonian is now strictly 2D.

Alternatively, we can argue as follows. Since
$k_F^{-1}\gg w_z$, we can approximate $|\mathbf r_i-\mathbf r_j|\approx |\mathbf r_{i//}-\mathbf r_{j//}|$
and therefore
\bea
&&\frac{1}{N_z}\sum\limits_{z_i,z_j=1}^{N_z} J^{\al\beta}(|\mathbf r_i-\mathbf r_j|) I_i^\alpha  I_j^\beta \approx
\left(\frac{1}{\sqrt{N_z}}\sum\limits_{z_i} I_i^\alpha(\mathbf r_{i//},z_i)\right) \nn\\
&& \times J^{\al\beta}(|\mathbf r_{i//}-\mathbf r_{j//}|)
\left( \frac{1}{\sqrt{N_z}}\sum\limits_{z_j} I_j^\beta(\mathbf r_{j//},z_j)\right).
\eea 
This amounts to replacing a single nuclear spin at position $(\mathbf r_{j//},z_j)$ 
by an average nuclear spin $\frac{1}{\sqrt{N_z}} \sum\limits_{z_j} \mathbf I_j(\mathbf r_{j//},z_j)$.
This is fully consistent with the fact that the RKKY interaction is almost constant and ferromagnetic
in the $z$-direction perpendicular to the 2DEG (this is at least the case for the RKKY interaction
obtained by neglecting electron-electron interactions but we expect that 
the interactions do not  modify
significantly the RKKY interactions in the $z$-direction).

Our problem has now been reduced to a 2D system consisting of $N=N_l/N_z$ nuclear spins
interacting with long-ranged interactions. The effective nuclear spin Hamiltonian $H_{\rm eff}=\la U\ra$
is finally given by Eq. (\ref{eq:ueff}).


\begin{thebibliography}{30}
\bibitem{note_slope} A visual inspection of the curvature at the origin of the 
QMC data in Refs. \onlinecite{qmc,GV} suggests $d\chi_s/dq(q=0)<0$ (case (c) in Fig. \ref{Fig:chis}) 
for small $r_s$,
and  $d\chi_s/dq(q=0)>0$ at larger $r_s\gtrsim 5$ (case (a) or (b) in Fig. \ref{Fig:chis}). 
If this is correct, this would imply that there is a finite value 
of $r_s>1$ where $d\chi_s/dq(q=0)=0$
as in the non-interacting case.
\bibitem{note_cq} In the RPTA, $\omega_q\approx c q/\ln^2(v_Fq/k_BT_L)$ at large q, 
but at low $q\lesssim \Delta/v_F\sim k_F$,  $\omega_q\approx c q/\ln^2(\Delta/k_BT_L)$ 
(see Eq. \ref{eq:gamma_cutoff}) and is therefore
linear in $q$ in the range of interest for spin waves.
\bibitem{note_cons} The $J^{\alpha\beta}_{ij}$ are proportional to the spin susceptibilities 
$\chi^{\alpha\beta}_{ij} \propto \int_0^\infty dt [ S_i^\alpha(t),S_j^\beta]$
(see Eq. (\ref{eq:susc})). If the total  spin of the electron system is conserved, we have
$\chi^{xx}_{ij} = \chi^{yy}_{ij} \propto 
\int_0^\infty dt ([ S_i^+(t),S_j^-] + [ S_i^-(t),S_j^+])$,
and 
$\chi^{xy}_{ij} = - \chi^{yx}_{ij} \propto 
i \int_0^\infty dt ([ S_i^+(t),S_j^-] - [ S_i^-(t),S_j^+])$,
with $S^\pm_i = S^x_i \pm i S^y_i$.
Furthermore, if the electron system is time-reversal invariant, we expect that
 $\chi^{\alpha\beta}_{ij} = 0$ for $\alpha \neq \beta$ because it allows
us to interchange the roles of the electron spins $\uparrow$ and $\downarrow$.
By the same reasoning, we find $J^{zz}_{ij} = J^{xx}_{ij}$
as well.
In order to maintain a bit of more generality, however, we keep $J^{zz}_{ij}$ here
distinct from $J^{xx}_{ij}$ and $J^{yy}_{ij}$.
\bibitem{heli_nonanalytic}
We assume for the analysis of Eq.~(\ref{eq:m_q0}) that $\chi_s(q)$
remains analytic in the vicinity of its minimum at $q_1$, and 
therefore $\chi_s(q) \propto (q-q_1)^2$.
If, however, $\chi_s(q)$ has a cusp at $q_1$, we have
$\chi_s(q) \propto |q-q_1|$ with two different slopes for
$q>q_1$ and $q<q_1$ respectively. The integral in Eq. (\ref{eq:m_q0}) is
no longer divergent, and the discussion of the transition temperature
follows along the lines of Sec. \ref{sec:fm}.
\end{thebibliography}
\end{document}